\documentclass[sn-mathphys,Numbered]{sn-jnl}

\usepackage{graphicx}
\usepackage{multirow}
\usepackage{amsmath,amssymb,amsfonts}
\usepackage{amsthm}
\usepackage{mathrsfs}
\usepackage[title]{appendix}
\usepackage[dvipsnames]{xcolor}
\usepackage{textcomp}
\usepackage{manyfoot}
\usepackage{booktabs}
\usepackage{algorithm}
\usepackage{algorithmicx}
\usepackage{algpseudocode}
\usepackage{listings}

\usepackage{bm, mathtools, url, hyperref, cancel, color}
\usepackage{subcaption}
\usepackage{relsize}

\newcommand{\T}{^{\mbox{\tiny T}}}
\newcommand{\B}[1]{{\bm #1}}
\newcommand{\U}[1]{\hat{\bm #1}}
\newcommand{\dd}{\; \text{d}}

\newcommand{\andd}{\qquad \text{and} \qquad}

\begin{document}

\title[Article Title]{Application of the Theory of Functional Connections to the Perturbed Lambert's Problem}

\author*[1]{\fnm{Franco} \sur{Criscola}}\email{criscolf@my.erau.edu}

\author[1]{\fnm{David} \sur{Canales}}\email{canaled4@erau.edu}

\author[2]{\fnm{Daniele} \sur{Mortari}}\email{mortari@tamu.edu}

\affil*[1]{\orgdiv{Aerospace Engineering}, \orgname{Embry-Riddle Aeronautical University}, \orgaddress{\street{1 Aerospace Boulevard}, \city{Daytona Beach}, \postcode{32114}, \state{FL}, \country{USA}}}

\affil[2]{\orgdiv{Aerospace Engineering}, \orgname{Texas A\&M University}, \orgaddress{\street{745 H.R. Bright Building}, \city{College Station}, \postcode{77843}, \state{TX}, \country{USA}}}

\abstract{A numerical approach to solve the perturbed Lambert's problem is presented. The proposed technique uses the Theory of Functional Connections, which allows the derivation of a constrained functional that analytically satisfies the boundary values of Lambert's problem. The propagation model is devised in terms of three new variables to mainly avoid the orbital frequency oscillation of Cartesian coordinates. Examples are provided to quantify robustness, efficiency, and accuracy on Earth and Sun centered orbits with various shapes and orientations. Differential corrections and a robust Lambert solver are used to validate the proposed approach in various scenarios and to compare it in terms of speed and robustness. Perturbations due to Earth's oblateness, third-body, and Solar radiation pressure are introduced, showing the algorithm's flexibility. Multi-revolution solutions are obtained. Finally, a polynomial analysis is conducted to show the dependence of convergence time on polynomial type and degree.}


\maketitle

\section{Introduction}\label{sec: intro}

\begin{figure}[b!]
	\centering
	\includegraphics[width=2.5in]{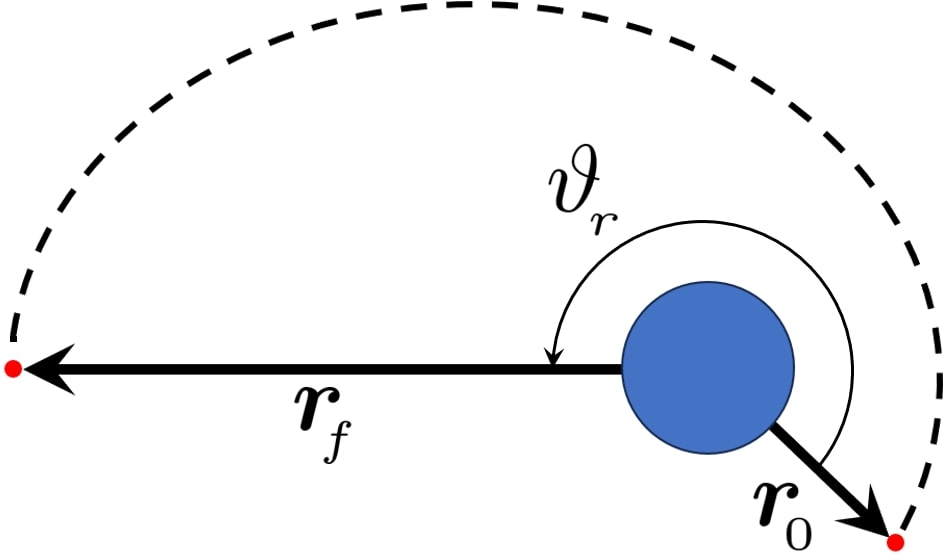}
	\caption{Lambert's problem geometry.}
	\label{fig: lambert}
\end{figure}

Lambert's problem is an important two-point boundary-value problem (BVP) in orbital mechanics. It is defined by the initial and final position of the spacecraft, as well as the time of flight (ToF) between them in an unperturbed scenario. The problem was first defined by Heinrich Lambert (1728-1777): \emph{In a two-body scenario with gravitational parameter of a planet, $\mu$, the time $\Delta T$ required to accomplish a given transfer is a function of (1) the semi-major axis $a$ of a conic trajectory joining the two position states, (2) the sum $\B{r}_0 + \B{r}_f$ of the distances from the primary at the beginning and at the end of the transfer, and (3) the linear distance c between the two points (chord)}. In other words, Lambert's problem leverages the system's dynamics and the boundaries to obtain the trajectory of a spacecraft. Figure~\ref{fig: lambert} shows the geometry of the problem, where $\B{r}_0$ and $\B{r}_f$ represent the initial and final positions, respectively, $\Delta T = t_f - t_0$, and $\vartheta_r$ is the angle covered by the arc connecting the boundaries.
This problem is commonly used in many applications, such as orbit determination, targeting, rendezvous~\cite{Fantino2021}, and in preliminary orbit design to build pork-chops. No closed-form solution to Lambert's problem is known, so authors have designed numerical algorithms that tend to trade robustness for efficiency, or \emph{vice versa}. Multiple solvers exist specially devised for unperturbed scenarios, with authors grouping around the selected free parameter~\cite{Lumpp1990, Peterson2010, Lee2016}, i. e., semi-major axis~\cite{Wailliez2014, Thorne1995, Thorne2014}, semi-latus rectum~\cite{Avanzini2008, Boltz1984}, or flight-path angle~\cite{Arlulkar2011}. 

There are several semi-analytical methods for solving the unperturbed Lambert's problem, which, though rooted in algebraic and geometric relations, necessitate some numerical computations, particularly root-finding algorithms. The first approach employs the universal variable algorithm, utilizing Lagrange coefficients and Stumpff functions to construct its algebraic solution, as detailed in Curtis' book~\cite{curtisbook}. The second method, the Lambert-Gauss algorithm, is based on Kepler's second law, emphasizing that equal areas are swept in equal times, and is thoroughly discussed in the literature~\cite{vallado2001fundamentals}. Lastly, Lagrange's method, which leverages equations from Lagrange's proof of Lambert's theorem and incorporates hypergeometric functions, is explained in Battin's book~\cite{battinbook}. Despite their algebraic foundations, these methods are semi-analytical, striking a balance between algebraic derivations and numerical accuracy, leading to faster computation times compared to purely numerical methods. Recent advancements in solving Lambert's problem have focused on enhancing the efficiency and accuracy of trajectory calculations. Lambert’s problem has been revisited with an unperturbed Lambert solver that significantly reduces computational complexity, converging in two iterations for single revolutions and three for multi-revolutions~\cite{Izzo2014}, distinctively utilizing Householder iterations for enhanced performance over Gooding’s algorithm~\cite{Gooding1990}. This method, inspired by Lancaster and Blanchard's work~\cite{Lancaster1968}, underscores the importance of initial guess and iteration methods to solve the problem. Russell's 2021 solver introduces a novel iterative approach using the vercosine formulation and interpolated initial guesses, remarkably reducing the total coefficient memory burden to just one megabyte and introducing a new invariant variable for problem reformulation~\cite{Russell2021}. Levi-Civita regularization has also been used~\cite{Fantino2018}, improving upon Carles Simo’s~\cite{Simo1973} method to achieve convergence on average in five iterations, illustrating the effectiveness of regularization in unperturbed Lambert solutions. 

These advanced methods share the aim of enhancing efficiency in the solving process by strategically refining initial guesses and iteration techniques using new invariant variables, typically designed for unperturbed scenarios, and building on the two-body problem (2BP) solution. However, it is worth noting that introducing perturbations into the model often results in increasing errors. Differential corrections, a prevalent root-finding technique, iteratively employs system dynamics to identify initial conditions meeting constraints in shooter algorithms~\cite{Canales2021}. Although this method is broadly used, it requires substantial previous work to find an initial guess that allows the algorithm to converge. In contrast to those previously mentioned, it allows for the introduction of multiple perturbations. Other solutions to the perturbed Lambert's problem exist, such as a method leveraging differential algebra optimization~\cite{Panicucci2018}. However, this method presents high computation times. One method developed to solve the $J_2$-perturbed Lambert's problem leverages the Kustaanheimo–Stiefel transformation and Modified Chebyshev–Picard iteration, enhancing the domain of convergence for Picard iteration in solving Keplerian problems~\cite{Woollands2015}. A key advantage of the proposed method is that it does not require a shooting method for single revolutions, and an initial guess for multi-revolution arcs is easily obtained from the unperturbed solution. On the other hand, deep neural networks have also been used to solve the perturbed Lambert's problem by determining an accurate initial guess~\cite{Yang2022}.

This research presents a comprehensive analysis on how to solve the perturbed Lambert's problem using the Theory of Functional Connections (TFC) ~\cite{TFC_Book}. Specifically, this article completes, by including methodology, results, and comparisons with other methods, some previous work presented by the authors ~\cite{Criscola2023,Criscola2023_2}. Some similarities may be found with different work from one of the authors~\cite{Johnston2018}. This formulation is different from such previous work, and shows significant improvement both in performance and flexibility, especially in solving multi-revolution problems. Here, new examples are introduced, together with an assessment of the nonlinear least-squares convergence adopted to estimate the solution, and an analysis of the orthogonal polynomial types to model the adopted solution variables. Thanks to TFC, the proposed nonlinear iterative least-squares approach does not need any algorithm to find an initial guess. The three new independent time-varying variables adopted for propagation model are $p (t)$, $\vartheta (t)$, and $h (t)$. These variables are devised to slightly deviate from constant and linear behaviors for small eccentric orbits and have specific known boundary values that are embedded by TFC into three corresponding constrained functionals. The three free functions of these functionals are then expressed using a set of orthogonal polynomials with unknown coefficients, which are solved using non-linear least squares. Because the constrained functionals adhere to the boundary conditions (BC) regardless of the free functions chosen, the unknown coefficients' initial guess is set to zero. This is because zero-valued free functions still produce solutions that fully meet the initial and final boundary requirements. Finally, the proposed formulation includes \textit{any} additional perturbations that are analytically expressed, such as a three-body, Solar radiation pressure (SRP), or $J_2$ perturbation~\cite{curtisbook}. For the sake of comparison, a differential corrector is used to highlight the advantages of the methodology proposed here. 

The paper is structured as follows. First, the mathematical model adopted to solve the perturbed Lambert's problem is introduced (Section~\ref{sec: math}). Then, multiple scenarios are used to generate unperturbed solutions (Section~\ref{sec: unpertSect}). A Differential Corrections (DC) algorithm and a robust solver that implements Izzo's~\cite{Izzo2014} and Blanchard's~\cite{Lancaster1968} algorithms are used to compare the performance of TFC. Then, three perturbations are implemented: Earth's oblateness, third-body, and Solar Radiation Pressure. Solutions are compared to the DC solver (Section~\ref{sec: pert}). Finally, a polynomial analysis is done by varying the degree and type via Gegenbauer polynomials (Section~\ref{sec: poly}). The three primary characteristics to be analyzed throughout this investigation are final position error (obtained trajectory with respect to expected), number of iterations, and computation times.

\section{Mathematical Model}
\label{sec: math}

The perturbed Lambert's problem consists of the following BVP:
\begin{equation}\label{PLP}
     \ddot{\B{r}} =-\dfrac{\mu}{r^3} \, \B{r} + \B{a}_p (\B{r}, \dot{\B{r}}) \qquad \text{subject to:} \quad \begin{cases} \B{r} (0) = \B{r}_0 \\ \B{r} (\Delta T) = \B{r}_f\end{cases}
\end{equation}
where $\B{a}_p$ indicates all external perturbing accelerations, $\B{r}_0$ and $\B{r}_f$ are the initial and final position vectors, and $\Delta T$ denotes the ToF (Figure~\ref{fig: lambert}). In this study, the TFC \cite{U-TFC,LDE,TFC_Book} is applied to solve this problem by representing the position vector, $\B{r} (t)$, with three new variables: the distance $p (t)$, the angle $\vartheta (t)$, and the out-of-plane coordinate $h (t)$. These variables identify the position vector as,
\begin{equation}\label{r}
    \B{r} (t) = p (t) \, \big[\U{r}_0 \, \cos\vartheta (t) + \U{t}_0 \, \sin\vartheta (t)\big] + h (t) \, \U{h}_0
\end{equation}
where $\U{r}_0 = \dfrac{\B{r}_0}{\mid \B{r}_0 \mid}$, $\U{r}_f = \dfrac{\B{r}_f}{\mid \B{r}_f \mid}$, $\U{h}_0 = \dfrac{\U{r}_0 \times \U{r}_f}{\mid \U{r}_0 \times \U{r}_f \mid}$, and $\U{t}_0 = \U{h}_0 \times \U{r}_0$. In particular, $[\U{r}_0, \U{t}_0, \U{h}_0]$ constitute the three directions of an orthogonal reference frame that is defined as long as the cross product $\U{r}_0 \times \U{r}_f$ exists, i.e., when Lambert's problems is not singular. The variable $p (t)$ represents the projection of the radius vector on the $[\U{r}_0, \U{t}_0]$ plane, $\vartheta (t)$ is a parametric angle with no physical meaning (other than satisfying the bounds), and $h (t)$ represents the orthogonal projection of the position vector with respect to the [$\U{r}_0, \U{t}_0$] plane. The first two derivatives of $\B{r}(t)$ are:
\begin{align}
    \label{rdot}
    \dot{\B{r}}(t) &= \big[\dot{p} \cos\vartheta - p \dot{\vartheta} \sin\vartheta\big] \hat{\bm r}_0 + \big[\dot{p} \sin\vartheta + p \dot{\vartheta} \cos\vartheta\big] \hat{\bm t}_0 + \dot{h} \, \hat{\bm h}_0 \\
    \nonumber
    \ddot{\B{r}}(t) &= \big[(\ddot{p} - p \dot{\vartheta}^2) \cos\vartheta - (2\dot{p} \dot{\vartheta} + p \ddot{\vartheta}) \sin\vartheta\big] \hat{\bm r}_0 + \\
    & \quad +\big[(\ddot{p} - p \dot{\vartheta}^2) \sin\vartheta + (2\dot{p} \dot{\vartheta} + p \ddot{\vartheta}) \cos\vartheta\big] \hat{\bm t}_0 + \ddot{h} \, \hat{\bm h}_0\label{rddot}
\end{align}
Consequently, the position, velocity, the acceleration vectors are expressed in terms of the variables, $p (t)$, $\vartheta (t)$, and $h (t)$. The boundary constraints of $\vartheta (t)$ and $h (t)$ are,
\begin{equation}\label{constraints}
     \begin{cases} \vartheta (0) = 0 \\ \vartheta (\Delta T) = \vartheta_r + 2k\pi\end{cases} \qquad \text{and} \qquad \begin{cases} h (0) = 0 \\ h (\Delta T) = 0\end{cases}
\end{equation}
where $\U{r}_0\T \U{r}_f = \cos \vartheta_r$ and $k$ is the number of complete revolutions. The boundary constraints of $p (t)$ are:
\begin{equation}\label{p}
     p (0) = \mid \B{r}_0 \mid = r_0 \andd p (\Delta T) = \mid \B{r}_f \mid = r_f
\end{equation}
TFC allows to derive constrained functionals, one for each variable, that always satisfy the boundary constraints given in Eqs. \eqref{constraints} and \eqref{p}. These constrained functional are (see \cite{TFC_Book} for the derivations),
\begin{equation}\label{rth}
     \begin{cases} p (t,g_p) = g_p (z) + \dfrac{\Delta T - t}{\Delta T} \big[p_0 - g_p (-1)\big] + \dfrac{t}{\Delta T} \big[p (\Delta T) - g_p (+1)\big] \\[8pt] \vartheta (t,g_{\vartheta}) = g_{\vartheta} (z) - \dfrac{\Delta T - t}{\Delta T} \, g_{\vartheta} (-1) + \dfrac{t}{\Delta T} \big[\vartheta_r + 2 k \pi - g_{\vartheta} (+1)\big] \\[8pt] h (t,g_h) = g_h (z) - \dfrac{\Delta T - t}{\Delta T} \, g_h (-1) - \dfrac{t}{\Delta T} \, g_h (+1)\end{cases}
\end{equation}
where $g_p (z)$, $g_{\vartheta} (z)$, and $g_h (z)$ are the three free functions associated with the three variables, $p (t)$, $\vartheta (t)$, and $h (t)$. It is straightforward to verify that the functionals in Eq. \eqref{rth} always satisfy the constraints, no matter what the free functions are. This is achieved by substituting \( t = 0 \) and \( t = \Delta T \) into Eq. \eqref{rth}, which produces functionals \( p, \theta, h \) at the boundaries. These functionals are then inserted into Eq.~\eqref{r}, demonstrating that the value of \(\mathbf{r}(t)\) satisfies the boundary conditions. Furthermore, these functional represent the whole set of functions satisfying the boundary constraints (see \cite{TFC_Book} for proof).

The free functions, $g_p (z)$, $g_{\vartheta} (z)$, and $g_h (z)$, are expanded in terms of a set of basis functions, $\B{s} (z)$,\footnote{Previous applications of TFC expanded the free functions in terms of Chebyshev or Lagrange orthogonal polynomials, which are defined in the range $z \in [-1, +1]$.} 
\begin{equation}\label{eq:gsDefinitions}
     g_p (z) = \B{\xi}_p\T  \B{s} (z), \qquad g_{\vartheta} (z) = \B{\xi}_{\vartheta}\T  \B{s} (z), \andd g_h (z) = \B{\xi}_h\T  \B{s} (z). 
\end{equation}
where $\B{\xi}_p$, $\B{\xi}_{\vartheta}$, and $\B{\xi}_h$ are the vectors of the unknown coefficients associated to $p (t)$, $\vartheta (t)$, and $h (t)$, respectively. Once these vectors are estimated, then the free functions are computed and the functionals given in Eq. \eqref{rth} provide the final estimated solution. Note that the solution provided is \textit{continuous}, that can be evaluated at any time, in contrast to solutions based on numerical integration where the solution is provided at specific integration times. In addition, a solution provided in terms of polynomials makes trivial subsequent derivative and/or integral manipulations. The derivatives of the free functions are,
\begin{equation}
     \begin{cases}\dot{g}_p = \B{\xi}_p\T \dot{\B{s}}, \qquad 
     \dot{g}_{\vartheta} = \B{\xi}_{\vartheta}\T \dot{\B{s}}, \qquad
     \dot{g}_h = \B{\xi}_h\T  \dot{\B{s}} \\
     \ddot{g}_p = \B{\xi}_p\T  \ddot{\B{s}}, \qquad \ddot{g}_{\vartheta} = \B{\xi}_{\vartheta}\T  \ddot{\B{s}} , \qquad
     \ddot{g}_h = \B{\xi}_h\T  \ddot{\B{s}}\end{cases}
\end{equation}
Additionally, the following approximate mean frequency
\begin{equation}
\label{eq: oscillation}
    \omega \approx \dfrac{2 k \pi + \vartheta_r}{\Delta T}
\end{equation}
captures the oscillation frequency of, $p (t)$, $\vartheta (t)$, and $h (t)$. This frequency information is included in the set of basis functions, $\B{s} (z)$, by adding trigonometry functions to the set of orthogonal polynomials. Therefore, the vector of the selected basis functions is expressed as,
\begin{equation}
\label{eq: basis function}
     \B{s} (z) = \begin{Bmatrix} \B{L} (z) \\ \cos(\omega \, t) \\  \sin(\omega \, t)\end{Bmatrix}\T
\end{equation}
where $\B{L} (z)$ is the set of orthogonal polynomials adopted (e.g., Legendre or Chebyshev). The Legendre and Chebyshev orthogonal polynomials are defined in the $z\in[-1,+1]$ range. Due to the use of orthogonal polynomials, the polynomials variable ranges between $[-1, +1]$. This implies mapping the time variable $t \in [0, \Delta T]$ with the polynomials variable and modifying all derivatives accordingly. Therefore, a linear mapping is introduced between $z$ and the time $t$:
\begin{equation}
    z (t) = \dfrac{2}{\Delta T} \, t - 1 \in [-1, +1] \qquad \to \qquad t (z) = \dfrac{\Delta T}{2} (z + 1) \in [0, \Delta T]
\end{equation}
This linear map allows the mapping of derivatives in $z$ and in $t$ as follows:
\begin{equation}
    \dot{g} = \dfrac{\dd g}{\dd t} = \dfrac{\dd g}{\dd z} \cdot \dfrac{\dd z}{\dd t} = g' \, c = g' \, \dfrac{2}{\Delta T} \andd \dfrac{\dd^k g}{\dd t^k} = \dfrac{\dd^k g}{\dd z^k} \, c^k
\end{equation}
where $c=\dfrac{2}{\Delta T}$ is a mapping constant between time range and orthogonal polynomials range. This implies that the two time derivatives of the basis functions are,
\begin{equation}
     \dot{\B{s}} = \begin{Bmatrix} c \, \B{L(z)}' \\ -\omega \, \sin(\omega \, t) \\ \omega \, \cos(\omega \, t)\end{Bmatrix}\T  \andd 
     \ddot{\B{s}} = \begin{Bmatrix} c^2 \, \B{L(z)}'' \\ -\omega^2 \, \cos(\omega \, t) \\ -\omega^2 \, \sin(\omega \, t)\end{Bmatrix}\T 
\end{equation}
and the time derivatives of the functionals are,
\begin{equation}\label{rthdot}
     \begin{cases} \dot{p} (t,g_p) = c \, g_p' (z) - \dfrac{1}{\Delta T} \left(p_0 - g_p (-1)\right) + \dfrac{1}{\Delta T} \big(p(\Delta T) - g_p (+1)\big) \\[8pt] \dot{\vartheta} (t,g_{\vartheta}) = c \, g_{\vartheta}' (z) + \dfrac{1}{\Delta T} \, g_{\vartheta} (-1) + \dfrac{1}{\Delta T} \big(\vartheta_r + 2k\pi - g_{\vartheta} (+1)\big) \\[8pt] \dot{h} (t,g_h) = c \, g_h' (z) + \dfrac{1}{\Delta T} \, g_h (-1) - \dfrac{1}{\Delta T} \, g_h (+1)\end{cases}
\end{equation}
where $g_p' (z)$, $g_{\vartheta}' (z)$, and $g_h' (z)$, indicate the first derivatives with respect to $z$. The second derivatives are:
\begin{equation}\label{rthddot}
    \ddot{p} (t,g_p) = c^2 \, g_p'' (z), \qquad \ddot{\vartheta} (t,g_{\vartheta}) = c^2 \, g_{\vartheta}'' (z), \andd \ddot{h} (t,g_h) = c^2 \, g_h'' (z)
\end{equation}
To derive the constrained functionals, TFC has used a constant and linear support functions. Now, the three free functions to estimate by least-squares must be linearly independent from the support functions adopted to derive the constrained functionals. For this reason, the constant, $L_0 (z) = 1$, and the linear term, $L_1 (z) = z$, must be removed from the basis functions.

To solve by least-squares for $\B{\xi}_p$, $\B{\xi}_{\vartheta}$, and $\B{\xi}_h$, the constrained functionals are substituted into the propagation model, and then into the dynamics equation, which is discretized. The number of discretization points increases accuracy, but also increases convergence time. Therefore, a trade off value is selected to balance computational load and accuracy. The Chebyshev-Gauss-Lobatto (CGL) points distribution was selected, 
    \begin{align}
        z &= -\cos\left(\pi\frac{k-1}{n-1}\right) & k &= 1,2,...n
    \end{align}
where $n$ is the total number of CGL points. By increasing the number of points at the boundary values the CGL points distribution compensates the Runge effect and, consequently, allows the use of higher degree polynomials on least-squares and, therefore, obtaining better accuracy.

The dynamics equation is written as,
\begin{equation}\label{eq:loss}
    \mathcal{L} = \ddot{\B{r}} + \dfrac{\mu}{r^3} \, \B{r} - \B{a}_p (\B{r}, \dot{\B{r}}) = \B{0}
\end{equation}
Linearizing around an estimated ($k$-th) solution, 
\begin{equation}
    \label{eq:NLLS}
    \B{0} \approx \mathcal{L}_{k}+
    \begin{bmatrix} \dfrac{\partial \mathcal{L}}{\partial \B{\xi}_p}, & \dfrac{\partial \mathcal{L}_x}{\partial \B{\xi}_{\vartheta}}, & \dfrac{\partial \mathcal{L}_x}{\partial \B{\xi}_h}
    \end{bmatrix}_k
    \begin{Bmatrix}
    \B{\xi}_p \\  \B{\xi}_{\vartheta} \\  \B{\xi}_h
    \end{Bmatrix}_k 
    =
    \mathcal{L}_{k} + \mathcal{J}_{k} \,  \B{\xi}_k \\
\end{equation}
from which the nonlinear least-squares solution is solved via an iterative process:
\begin{equation}\label{eq:iter}
    \B{\xi}_{k+1} = \B{\xi}_k - \left(\mathcal{J}_k\T  \mathcal{J}_k\right)^{-1} \mathcal{J}_k\T  \mathcal{L}_k
\end{equation}
where $\B{\xi}_k\T =\{\B{\xi}_p\T, \; \B{\xi}_{\vartheta}\T, \; \B{\xi}_h\T\}_k$, and $\mathcal{J}_{k}$ is the Jacobian matrix of the system. The Jacobian requires the expression of the partial derivatives with respect to the three unknown vectors of coefficients (provided in Appendix~\ref{sec: appendixPartials}). Once the formulation is complete, the equations must be evaluated iteratively due to the indirect and direct dependence on $\B{\xi}_p$, $\B{\xi}_{\vartheta}$, and $\B{\xi}_h$. Finally, the initial guess for $\B{\xi}_{p 0}$, $\B{\xi}_{\vartheta 0}$, and $\B{\xi}_{h 0}$ to start the nonlinear iterative process may be improved if a prior approximate solution (e.g., the unperturbed solution) is available.
\begin{enumerate}
    \item If no prior knowledge is known then $\B{\xi}_{p 0} = \B{\xi}_{\vartheta 0} = \B{\xi}_{h 0} = \B{0}$. This means that the nonlinear iterative approach begins with an initial trajectory that linearly changes the values of $p (t)$, $\vartheta (t)$, and $h (t)$, from their initial to their final values. 
    \item If prior knowledge of the problem is known, and the ToF is desired to be varied or perturbations are added, then the initial values of $\B{\xi}_{p 0}$, $\B{\xi}_{\vartheta 0}$, and $\B{\xi}_{h 0}$ are obtained by best-fitting the initial known trajectory using the constrained functionals. This is done by obtaining the final $\B{\xi}_i$ from the unperturbed problem.  
\end{enumerate}
Once an initial guess for $\B{\xi}_p$, $\B{\xi}_{\vartheta}$, and $\B{\xi}_h$ is provided, the nonlinear least-squares problem is initiated. However, highly sensitive perturbed problems require a more accurate initial guess for the vector of coefficients used to solve the problem, $\B{\xi}$. For this investigation, these perturbed problems are solved using the unperturbed solution as the initial guess. Note that more complex scenarios, such as the circular restricted three-body problem, would necessitate an even better-estimated guess, but this case is not considered in this study.

\section{Analysis of the unperturbed Lambert problem\label{sec: unpertSect}}

Before including perturbations, this section validates some two-body scenarios in LEO and MEO by varying eccentricities and inclinations, as well as cislunar and interplanetary transfers. The objective of this section is to demonstrate the algorithm's capability to generate solutions with various conditions and parameters. Note that a thorough analysis is conducted to determine the polynomial order, ultimately selecting those most efficient. More information about the polynomial order dependence is provided in a later section. As mentioned in the previous section, the TFC solver is initialized with three null vectors each of size $m + 1$, where $m$ is the degree of the chosen polynomial. The convergence tolerance for all cases and algorithms is set to 10$^{-9}$. Note that all these cases are generated using non-dimensional units that force $\mu=1$, thus reducing the computational load. The number of CGL points is set to 200.

\subsection{Case scenarios}\label{sec: unpert}

\begin{figure}
    \centering
    \includegraphics[width=\textwidth]{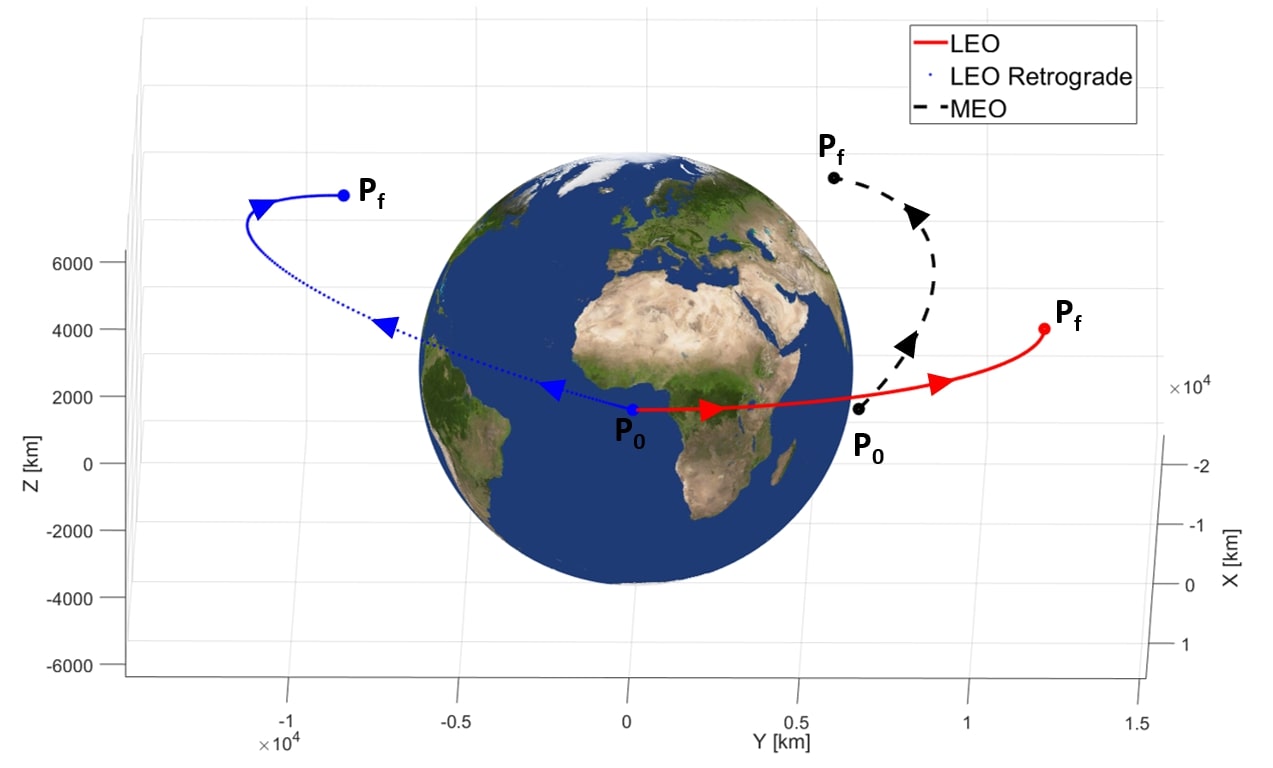}
    \caption{Two example orbits in LEO and MEO.}
    \label{fig: earthOrbits}
\end{figure}

\begin{figure}[h!]
    \centering
    \includegraphics[width=0.85\textwidth]{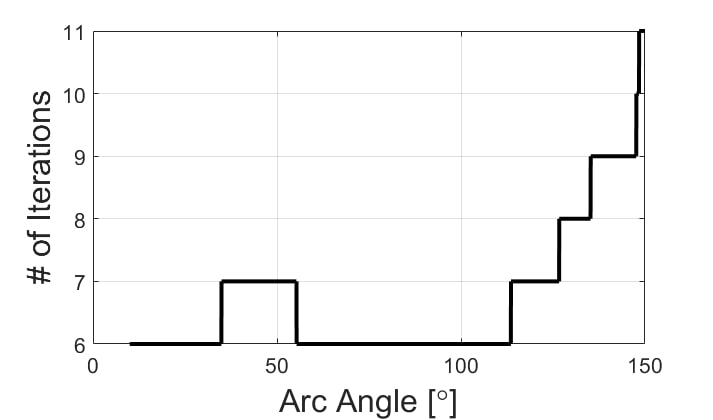}
    \caption{Number of iterations as a function of arc angle  for a MEO-GEO transfer.}
    \label{fig: angleVar}
\end{figure}

\begin{figure}[t]
    \centering
    \includegraphics[width=0.75\textwidth]{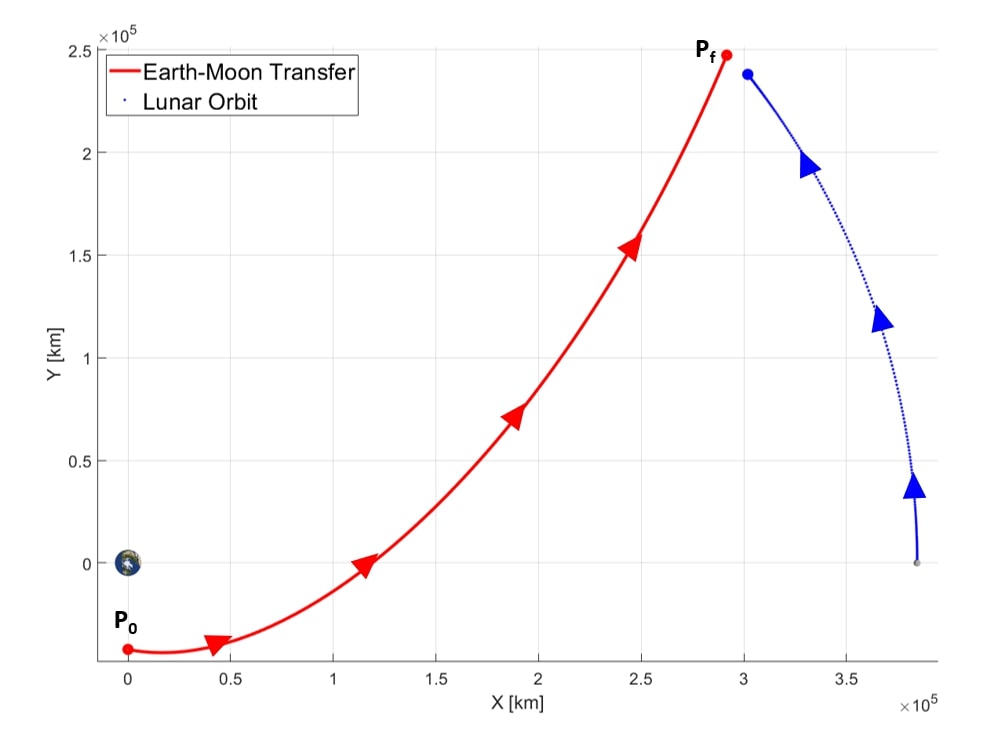}
    \caption{Cislunar} transfer arc.
    \label{fig: lunar}
\end{figure}

\begin{figure}[t]
    \centering
    \includegraphics[width=0.7\textwidth]{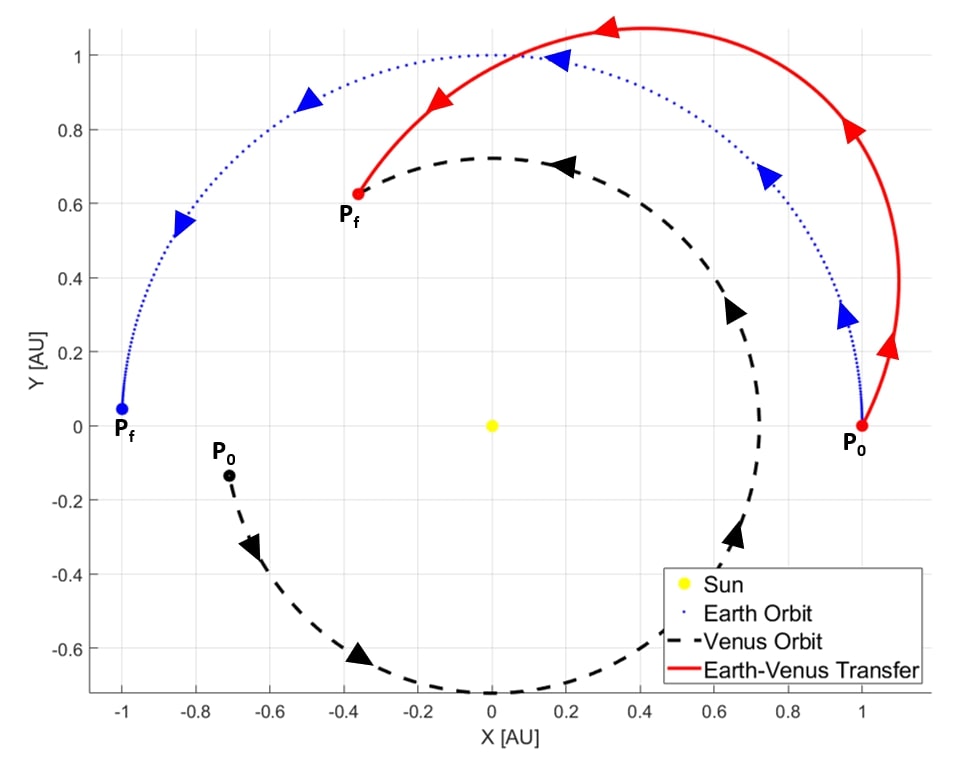}
    \caption{Sun-centered Earth-Venus transfer arc.}
    \label{fig: solar}
\end{figure}

Figure~\ref{fig: earthOrbits} shows three arcs (Table~\ref{tab: orbParams}) generated via TFC with Legendre polynomials of degree 15, given different periapsis altitude, eccentricity, inclination, and right ascension of the ascending node (RAAN). Figure~\ref{fig: angleVar} provides the number of iterations as a function of arc angle $\vartheta_r$ for a transfer from MEO ($2,000$ km altitude) to GEO for a fixed ToF (2.5 hours), showing that the number of iterations increases with the arc angle due to the singularity at 180$^\circ$. Other scenarios are generated in order to test the algorithm's ability to find solutions with larger distances, increased ToF, and independent on the primary body. Figure~\ref{fig: lunar} illustrates a cislunar 70-hours transfer arc with a $\vartheta_r$ = 130$^\circ$ angle, crafted using a 50th-degree Legendre polynomial. Note that the perturbation due to the Moon has not been included yet.
Finally, Figure~\ref{fig: solar} shows a Sun-centered Earth-Venus, 120$^\circ$, 180-day long transfer arc generated with a Legendre polynomial of degree 15. The results obtained in this section validate the robustness of the TFC approach to find solutions for various mission types. 

\begin{table}[b!]
    \fontsize{10}{10}\selectfont
    \centering 
    \caption{Orbital parameters of trajectories in Figure ~\ref{fig: earthOrbits}.}
    \begin{tabular}{ | c | c | c | c | } 
        \hline 
        \textbf{Parameter} & \textbf{LEO} & \textbf{LEO Retrograde} & \textbf{MEO} \\
        \hline
        Periapsis Altitude & 500 km & 500 km & 3,000 km \\
        \hline
        Eccentricity & 0.5 & 0.5 & 0 \\
        \hline
        Inclination & 0$^\circ$ & $165^\circ$ & $30^\circ$ \\
        \hline
        RAAN & $0^\circ$ & $0^\circ$ & $45^\circ$ \\
        \hline
        Arc Angle & $120^\circ$ & $150^\circ$ & $90^\circ$ \\
        \hline
        Polynomial Degree & 15 & 15 & 15 \\
        \hline
    \end{tabular}
    \label{tab: orbParams}
\end{table}


\subsection{Comparison with additional solvers}\label{sec:TFCDC}

Although Section~\ref{sec: unpert} shows that the obtained results are consistent with the two-body dynamics, it is also important to show that they are comparable in terms of performance with typical solvers, such as differential corrections and a Robust Lambert Solver (RLS). An analysis is thus provided here that compares the boundary condition's accuracy, the total computation time, and the number of iterations for all three algorithms. These performance criteria are analyzed under varying ToF, arc angle, and cord length.

\subsubsection{Differential corrections background}
The DC algorithm is a common iterative approach to solve boundary values problems, such as the Lambert's problem, with the purpose of minimizing the norm of the final error vector:
\begin{equation}
    \mathcal{F}(\mathcal{\B{Y}}) = \begin{Bmatrix} X_a - X_t, & Y_a - Y_t, & Z_a - Z_t\end{Bmatrix}\T
\end{equation}
where $\{X_a, Y_a, Z_a\}$ are the arrival coordinates obtained via numerical propagation, and $\B{r}_f = \{X_t, Y_t, Z_t\}\T$ is the final, desired position vector. Note that, in this problem, $\mathcal{\B{Y}}$ contains the velocity components at $\B{r}_0$:
\begin{equation}
    \mathcal{\B{Y}} = \begin{Bmatrix} \dot{X}_0, & \dot{Y}_0, & \dot{Z}_0\end{Bmatrix}\T
\end{equation}
The constraint vector is computed at each iteration until the Euclidean norm approaches an arbitrary tolerance value. If the condition is not met, the vector of free variables is updated as follows:
\begin{equation}
    \mathcal{\B{Y}} = \mathcal{\B{Y}}_0 - \mathcal{J}(\mathcal{\B{Y}}_0)^{-1}\mathcal{F}(\mathcal{\B{Y}})
\end{equation}
where $\mathcal{\B{Y}}$ is the updated initial velocity vector, $\mathcal{\B{Y}}_0$ is the current velocity vector, and $\mathcal{J}(\mathcal{\B{Y}}_0)$ is the Jacobian, easily retrieved from the state transition matrix:
\begin{equation}
    \mathcal{J}(\mathcal{\B{Y}}_0) =
    \begin{bmatrix}
        \dfrac{\partial X_a}{\partial\dot{X}_0} & \dfrac{\partial X_a}{\partial\dot{Y}_0} & \dfrac{\partial X_a}{\partial\dot{Z}_0} \\[5pt]
        \dfrac{\partial Y_a}{\partial\dot{X}_0} & \dfrac{\partial Y_a}{\partial\dot{Y}_0} & \dfrac{\partial Y_a}{\partial\dot{Z}_0} \\[5pt]
        \dfrac{\partial Z_a}{\partial\dot{X}_0} & \dfrac{\partial Z_a}{\partial\dot{Y}_0} & \dfrac{\partial Z_a}{\partial\dot{Z}_0}
    \end{bmatrix}
\end{equation}
This formulation concludes the basis of the DC algorithm.

\subsubsection{Performance analysis}

\begin{figure}[t!]
\centering
\begin{subfigure}{.33\textwidth}
  \centering
  \includegraphics[width=1\linewidth]{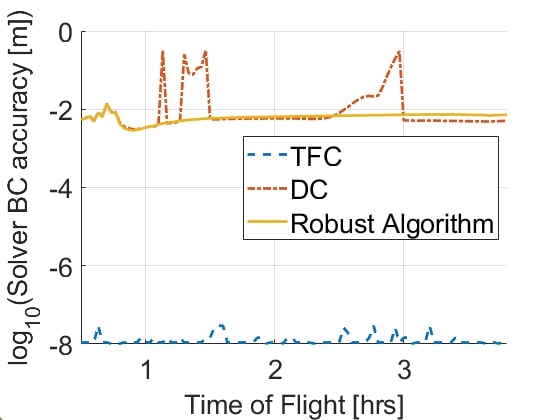}
  \caption{\label{TOF1}Final BC error vs ToF}
\end{subfigure}%
\begin{subfigure}{.33\textwidth}
  \centering
  \includegraphics[width=1\linewidth]{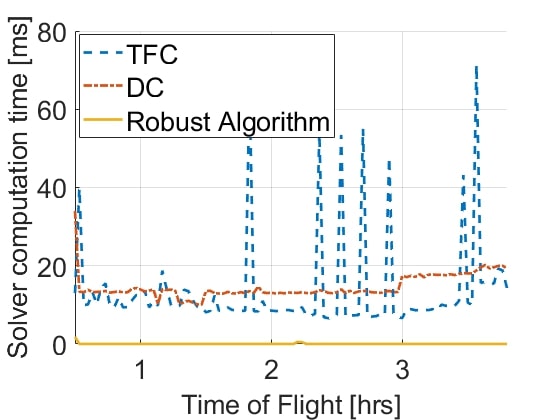}
  \caption{\label{TOF2}Computation time vs ToF}
\end{subfigure}
\begin{subfigure}{.33\textwidth}
  \centering
  \includegraphics[width=1\linewidth]{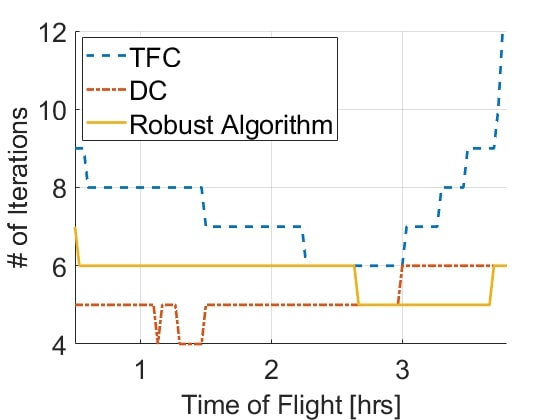}
  \caption{\label{TOF3}Iterations vs ToF}
\end{subfigure}
\begin{subfigure}{.33\textwidth}
  \centering
  \includegraphics[width=1\linewidth]{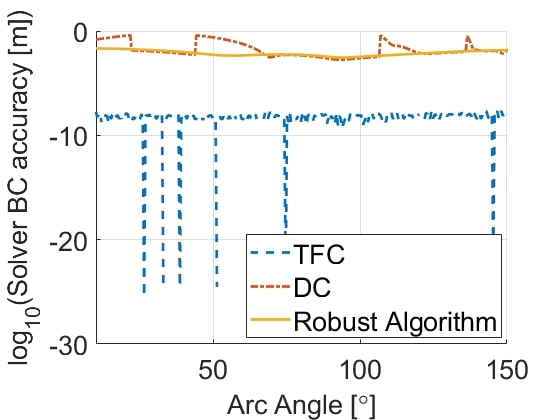}
  \caption{\label{ANG1}Final BC error vs $\vartheta_r$}
\end{subfigure}%
\begin{subfigure}{.33\textwidth}
  \centering
  \includegraphics[width=1\linewidth]{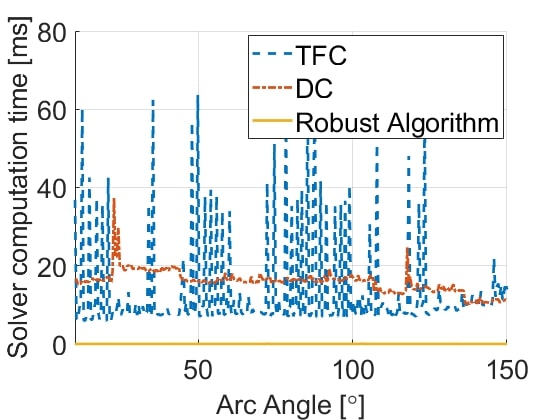}
  \caption{\label{ANG2}Computation time vs $\vartheta_r$}
\end{subfigure}
\begin{subfigure}{.33\textwidth}
  \centering
  \includegraphics[width=1\linewidth]{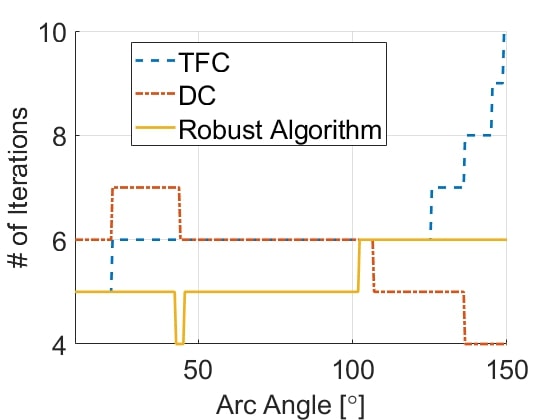}
  \caption{\label{ANG3}Iterations vs $\vartheta_r$}
\end{subfigure}
\begin{subfigure}{.33\textwidth}
  \centering
  \includegraphics[width=1\linewidth]{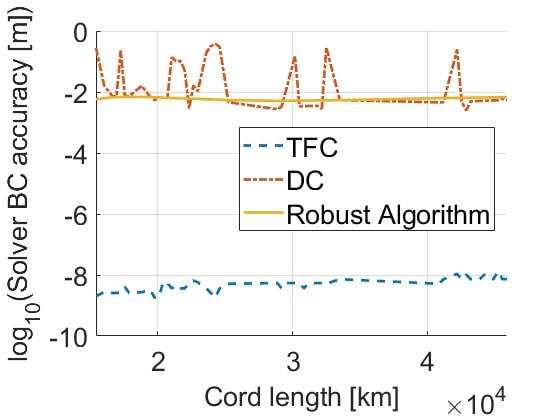}
  \caption{\label{CORD1}Final BC error vs cord}
\end{subfigure}%
\begin{subfigure}{.33\textwidth}
  \centering
  \includegraphics[width=1\linewidth]{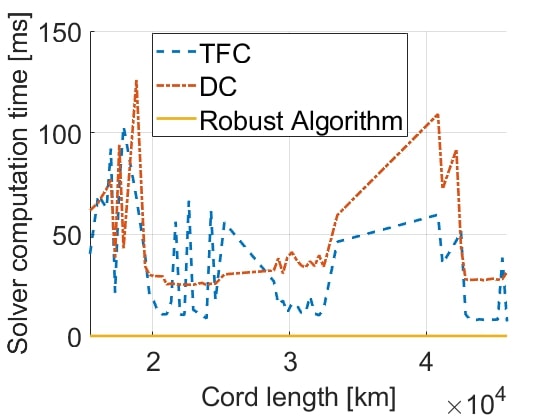}
  \caption{\label{CORD2}Computation time vs cord}
\end{subfigure}
\begin{subfigure}{.33\textwidth}
  \centering
  \includegraphics[width=1\linewidth]{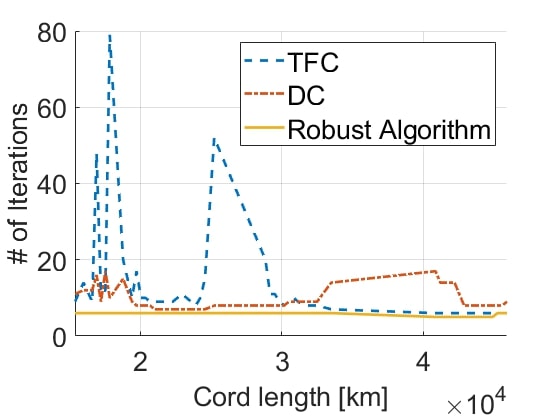}
  \caption{\label{CORD3}Iterations vs cord}
\end{subfigure}
\caption{\label{fig:performanceAnalysis}Analysis of performance of TFC, DC and RLS for different Lambert arcs.}
\end{figure}

All three algorithms (TFC, DC, and RLS) are now used to generate solutions for a MEO-GEO transfer arc with varying ToF, arc angle, and cord length (Figure \ref{fig:performanceAnalysis}). The DC method is initialized in this investigation using a Hohmann transfer as the initial guess, while the RLS does not require an initial guess. TFC does not need a pre-calculated initial guess like DC and it does not need to include the velocity components; TFC finds the optimal arc that links two boundary conditions in a given ToF considering the dynamics of the model. A polynomial degree of $20$ is used. Figures~\ref{TOF1} - \ref{TOF3} compare the performance of all three algorithms for two fixed boundary conditions solved for a given TOF between $0.5$ to $3.8$ hours. The spacecraft starts at an altitude of $1,500$ km and ends at GEO, with an arc angle of $120^\circ$. These results show that TFC is superior to the other algorithms in terms of accuracy, and inferior in terms of number of iterations for larger transfer angles. Figure~\ref{TOF2} leads to some interesting conclusions; given the nature of the RLS, it is expected that this method converges faster than numerical algorithms like TFC and DC. Comparing TFC and DC shows that the effect of Lambert problem's singularity as well as good initial guess requirement for DC have a significant effect on computation time. This is primarily due to the presence of matrix inverse operations.
Figures~\ref{ANG1}-\ref{ANG3} display the results of varying $\vartheta_r$ from $10^\circ$ to $150^\circ$ while fixing the ToF to $2.5$ hours and fixing the cord. Similar conclusions as before are drawn from these results: (1) the superiority of performance of TFC to meet the boundary conditions, and (2) the effect of DC's initial guess is more prominent here. Finally, Figures~\ref{CORD1}-\ref{CORD3} show the results of varying the cord length of the arc, i.e., changing the distance between the two BC. The arc angle is fixed to $120^\circ$ and the ToF to $2.5$ hours. The cord length varies from $15,406$ km to $45,906$ km, representing a final BC from MEO to GEO. Similar conclusions are drawn from these plots compared to the ones above ones. However, a potential problem with TFC is apparent in Figures~\ref{CORD2} and~\ref{CORD3}. The prominent peaks in both computation time and iteration count shows some divergence conditions under the given cord length. Further investigation shows that varying the polynomial type and degree eliminates these outliers, and also decreases TFC's computation time (Section~\ref{sec: poly}). Nonetheless, given that no pattern is easily identified, the results from this figure are inconclusive. Although the number of iterations and computation time increase in some cases, the accuracy of the final position error is unaffected, due to TFC's formulation. To conclude, it is apparent that TFC is not comparable in speed and iteration number with analytical solvers. However, due to Eqs.~\eqref{r} and ~\eqref{rth}, it is able to guarantee the BC to absolute and analytical precision. Similarly, determining the polynomial coefficients via least-squares allows the determination of the positions at the CGL points without the need of integration, thus reducing error.

\section{Perturbed problems}\label{sec: pert}

The ability to consider alternate forces outside of the primary's gravitational attraction is a key feature of solving Lambert's problem using TFC. In this analysis, the perturbations are included separately, but they may be added together within the context of the formulation. Firstly,  resultant perturbed orbits are compared to the unperturbed orbits generated via TFC. Then, the perturbed orbits are compared to a DC algorithm to determine TFC's ability to find  accurate solutions. This section also contains performance comparison as in the case of the unperturbed problem. As before, the convergence tolerance is set to 10$^{-9}$ and non-dimensional quantities are used. The number of CGL points is set to $200$. Note that the perturbed algorithms are initialized with an approximate solution obtained from the converged unperturbed problem.

\subsection{Perturbation models}

Three distinct perturbing accelerations are considered and tested. The Earth's oblateness ($J_2$), the solar radiation pressure, and third-body perturbations. The Jacobian is then built with the partial derivatives of $\B{a}_{J_2}$, $\B{a}_{3b}$, and $\B{a}_{_{\rm SRP}}$ with respect to $\B{r}$ (provided in Appendix~\ref{sec: appendixPartials}). Note that the Jacobian requires the conversion from $[\U{r}_0, \U{t}_0, \U{h}_0]$ into the inertial $[\U{x}, \U{y}, \U{z}]$ frame. 
\begin{itemize}
\item \textbf{Earth's oblateness.} The $J_2$ perturbation is expressed as \cite{curtisbook}: 
\begin{equation}
    \label{eq:acc_J2}
   \B{a}_{J_2} = -\dfrac{3 \, J_2 \, \mu \, r_{\rm eq}^2}{2} \, \dfrac{1}{r^5} \begin{Bmatrix} x \big[1 - 5 (z/r)^2\big] \\[8pt] y \big[1 - 5 (z/r)^2\big] \\[8pt] z \big[3 - 5 (z/r)^2\big]\end{Bmatrix}
\end{equation}
where $J_2 = 1.082629\cdot 10^{-3}$ and $r_{\rm eq} = 6,378.137$ km is the Earth's equatorial radius.

\item \textbf{Third-body.} The third-body perturbation is expressed as \cite{vallado2001fundamentals}:
\begin{equation}
   \label{eq:acc_3b}
   \B{a}_{3b} = \mu_{3b}\left( \dfrac{\B{r}_{sc-3b}}{{r}_{sc-3b}^3}-\dfrac{\B{r}_{3b}}{{r}_{3b}^3} \right)
\end{equation}
where $\mu_{3b}$ is the gravitational parameter of the third-body, $\B{r}_{sc-3b}$ is the vector of the third-body relative to the spacecraft, and $\B{r}_{3b}$ is the position vector of the third-body relative to the primary, all in inertial frame.

\item \textbf{Solar radiation pressure.} The SRP perturbation is expressed as~\cite{zardain}:
\begin{equation}
    \label{eq:acc_srp}
    \B{a}_{_{\rm SRP}} = \frac{P_{_{\rm SRP}} A}{m} \left[\rho_a \, (\U{n}\T \U{r}_s) \U{r}_s + 2\rho_s \, (\U{n}\T \U{r}_s)^2 \U{n} + \rho_d \, (\U{n}\T \U{r}_s) \left(\U{r}_s + \dfrac{2}{3} \U{n}\right)\right]
\end{equation}
where $\rho_a$, $\rho_s$, and $\rho_d$ are the relative material reflectivity properties of the surface ($\rho_a + \rho_s + \rho_d = 1$), $A$ is the surface area exposed to the Sun, $\U{n}$ the normal direction to the surface, $\U{r}_s$ the Sun-to-satellite direction, and $P_{_{\rm SRP}}$ is the SRP, given by $4.57\cdot 10^{-6}$ N/m$^2$.
\end{itemize}

\subsubsection{Earth's oblateness}

\begin{figure}[t!]
    \centering
    \begin{subfigure}{0.45\textwidth}
        \includegraphics[width=1.05\textwidth]{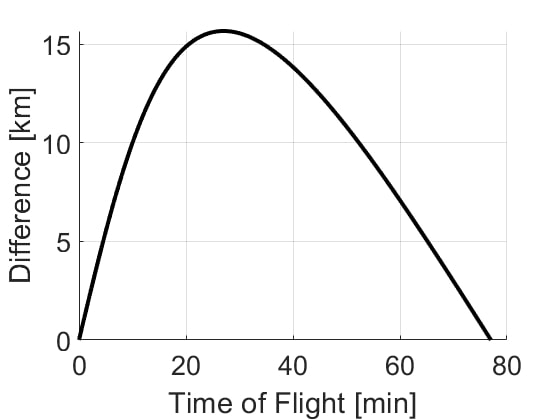}
        \caption{$J_2$-perturbed vs unperturbed Lambert arcs.}
        \label{fig: J2PertUnpertError}
    \end{subfigure}
    \begin{subfigure}{0.45\textwidth}
        \includegraphics[width=1.05\textwidth]{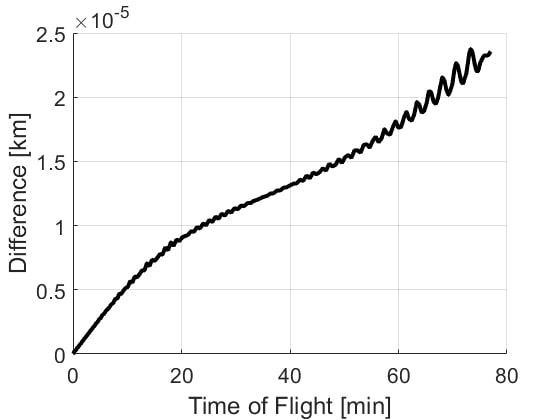}
        \caption{TFC vs differential corrections algorithms.}
        \label{fig: J2PertDiffCorrError}
    \end{subfigure}
    \caption{Error plots showing the norm of the vector difference between each state.}
    \label{fig: J2error}
\end{figure}

Orbits near Earth's surface are subject to the non-symmetric Earth's gravitational field. Although minor in small time scales, these effects stack over time, severely affecting the resulting orbit. This perturbation directly affects the RAAN and argument of periapsis depending on its inclination, altitude, and eccentricity~\cite{curtisbook}. A 150$^\circ$, 77-min long arc is generated with a Legendre polynomial degree of $50$ to test TFC's ability to find solutions under the effects of this perturbation. The initial semi-major axis of this orbit is $13,316$ km, the eccentricity is $0.5$, and the inclination is $50^\circ$. The choice of orbital parameters is meant to increase the effects of the perturbation. Although the magnitude of the perturbation is too small to visually notice in the duration of the transfer, Figure~\ref{fig: J2PertUnpertError} shows that the perturbation exists. The $y$-axis corresponds to the difference between the perturbed and unperturbed solutions. Figure~\ref{fig: J2PertDiffCorrError} shows the difference between TFC and DC. The error is minimal, proving that TFC is capable of obtaining effective solutions.

\begin{figure}[t!]
    \centering
    \begin{subfigure}{0.45\textwidth}
        \includegraphics[width=1\textwidth]{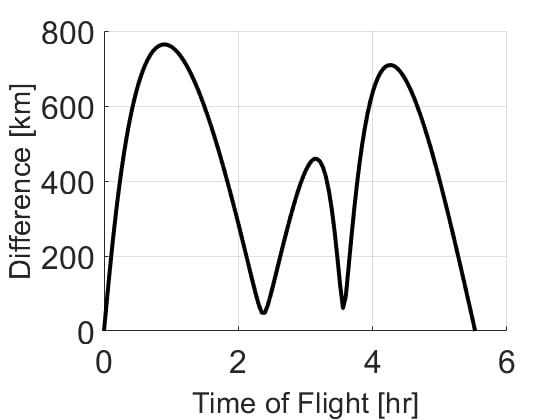}
        \caption{Multi-revolution perturbed vs unperturbed trajectories.}
        \label{fig: MultiRevTraj}
    \end{subfigure}
    \begin{subfigure}{0.5\textwidth}
        \includegraphics[width=1\textwidth]{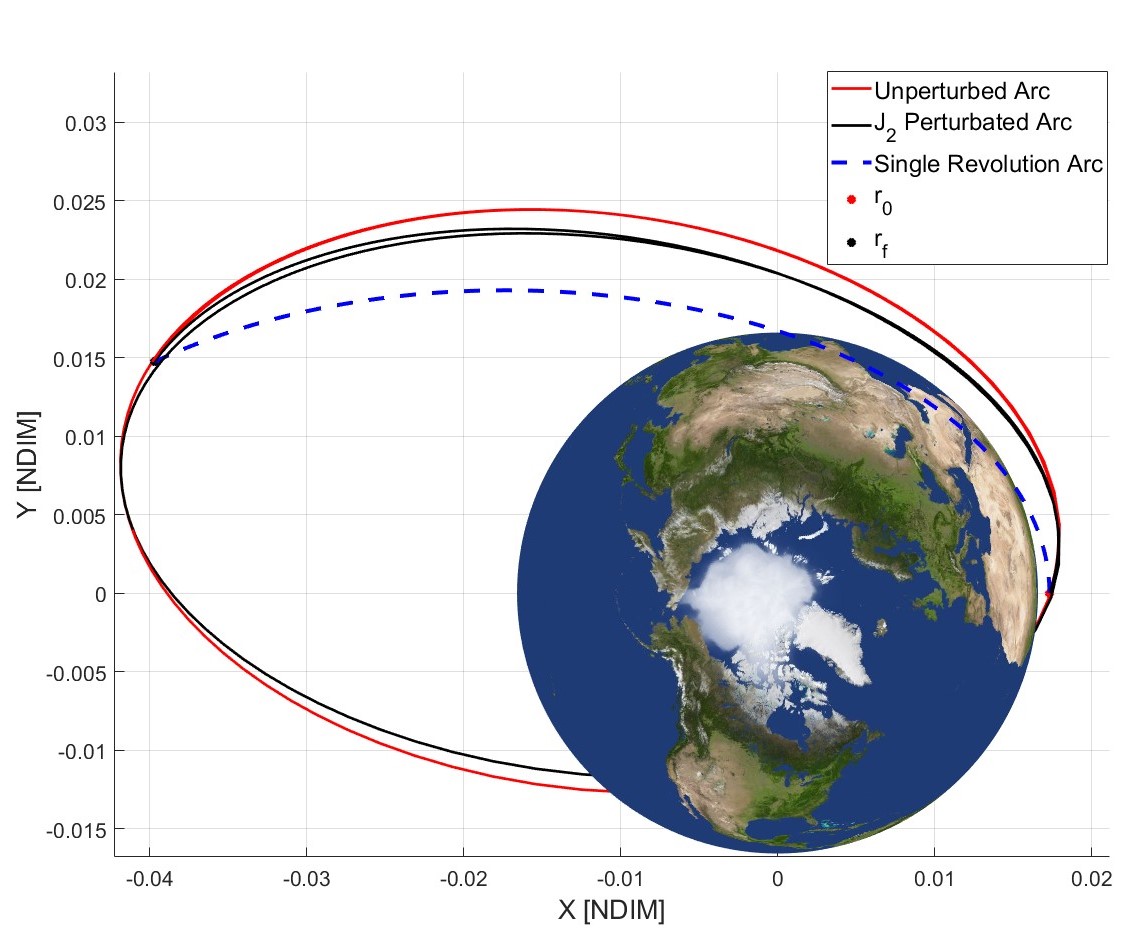}
        \caption{Top-down view of a multi-revolution Lambert arc generated via TFC.}
        \label{fig: MultiRev}
    \end{subfigure}
    \caption{Error plot of a multi-revolution solution trajectory.}
    \label{fig: MultiRevolution}
\end{figure}

Due to the presence of oscillatory terms in Eq.~\ref{r}, TFC is able to generate multi-revolution solutions. In unperturbed scenarios, this kind of solution is redundant, given that there is only one ToF that is able to find a solution. However, if a perturbation is added, the solutions become slightly more complex. In this investigation, multi-revolution solutions are found using the aforementioned $J_2$ perturbation. To exaggerate the effects of this perturbing force, the $J_2$ coefficient is increased by a factor of $10$ (Figure~\ref{fig: MultiRevolution}). The Legendre polynomial in this multi-revolution case increased from $50$ to $91$; this comes from the need for the polynomial to represent multiple revolutions of the orbit. The figure is showing the difference between an unperturbed and a perturbed multi-revolution arcs. The periodicity of the orbit is clear from the rise and fall of the error.

\subsubsection{Third-body perturbations}

\begin{figure}[b!]
    \centering
    \includegraphics[width=0.8\textwidth]{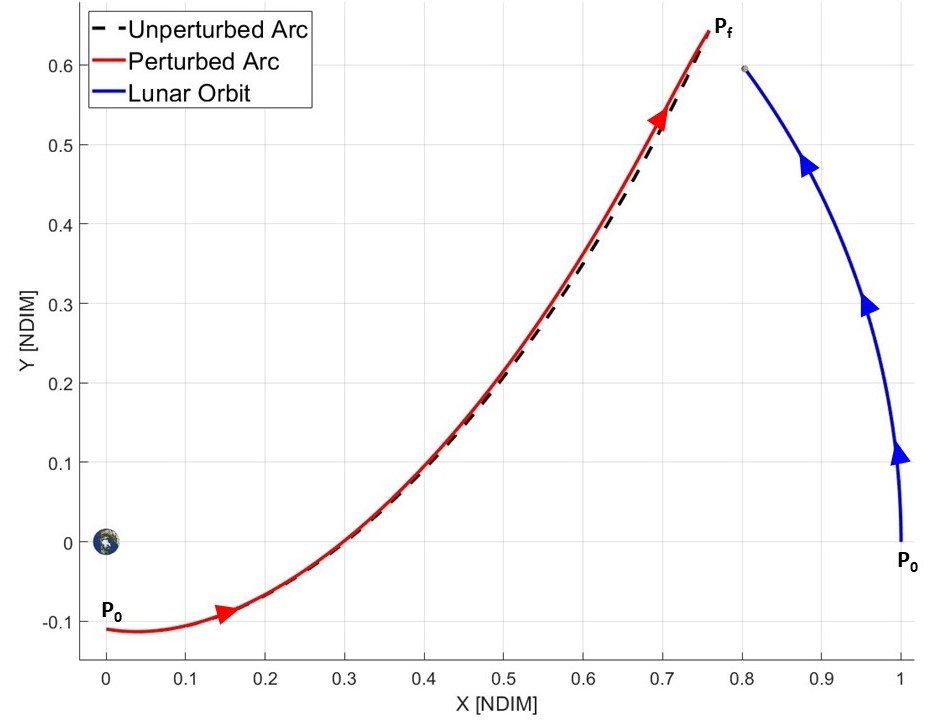}
    \caption{Perturbed Lambert arc under third-body perturbation.}
    \label{fig: 3BPOrbit}
\end{figure}

\begin{figure}[t!]
    \centering
    \begin{subfigure}{0.45\textwidth}
        \includegraphics[width=1\textwidth]{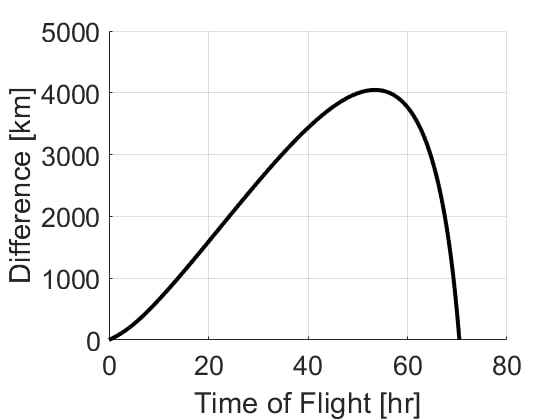}
        \caption{Third-body perturbed vs unperturbed Lambert arcs.}
        \label{fig: 3BPPertUnpertError}
    \end{subfigure}
    \begin{subfigure}{0.45\textwidth}
        \includegraphics[width=1\textwidth]{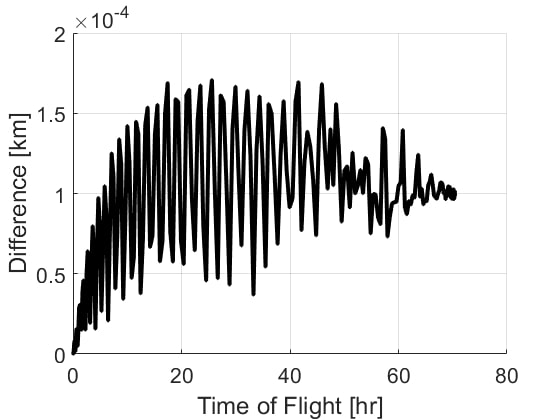}
        \caption{TFC vs DC algorithms.}
        \label{fig: 3BPPertDiffCorrError}
    \end{subfigure}
    \caption{Error plots showing the norm of the vector difference between each state.}
    \label{fig: 3BPerror}
\end{figure}

Solving Lambert's problem while considering the effects of the Moon's gravity is of particular importance as the cislunar region increases in popularity. The TFC framework allows for the inclusion of any number of perturbing bodies. In this investigation, the effect of the Moon is included (Eq.~\ref{eq:acc_3b}). A $130^\circ$ arc is generated with a Legendre polynomial degree of 50 to find a transfer from $\B{r}_0 = \{0; \; -42,164; \; 0\}\T$ km to $\B{r}_f = \{291,644; \; 247,332; \; 0\}\T$ km over a ToF of $70$ hours, reaching a distance of approximately $13,000$ km from the Moon. The Moon is propagated in conjunction with the spacecraft, starting at position $\{384,000; \; 0; \; 0\}\T$ km. Note that the magnitude of the perturbation is dependent upon the initial state of the Moon at departure. Figure~\ref{fig: 3BPOrbit} shows the results of the simulation, where the dotted line is the unperturbed Lambert arc, the solid line is the perturbed arc, and the right-most line is the orbit of the Moon. Figure~\ref{fig: 3BPPertUnpertError} shows the difference between the perturbed and unperturbed arcs arcs. Figure~\ref{fig: 3BPPertDiffCorrError} shows the difference between TFC and DC. As expected, the first plot shows a larger difference close to the Moon, as its gravitational influence increases. Most importantly, note that the boundary condition is still met at the end of the transfer. The second plot shows no significant difference between the TFC and DC solutions.

\subsubsection{Solar radiation pressure}

\begin{figure}[t!]
    \centering
    \includegraphics[width=0.85\textwidth]{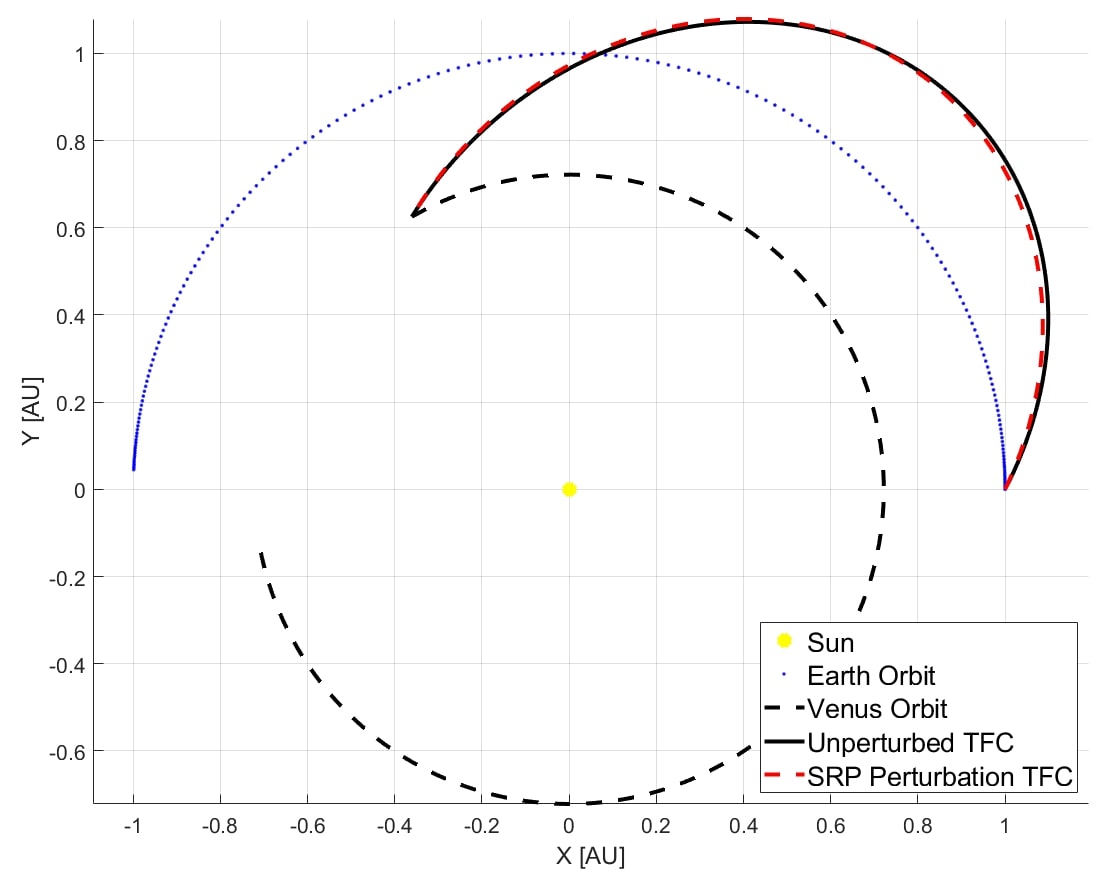}
    \caption{Perturbed Lambert arc under Solar radiation pressure.}
    \label{fig: SRPOrbit}
\end{figure}

\begin{figure}
    \centering
    \begin{subfigure}{0.45\textwidth}
        \includegraphics[width=1\textwidth]{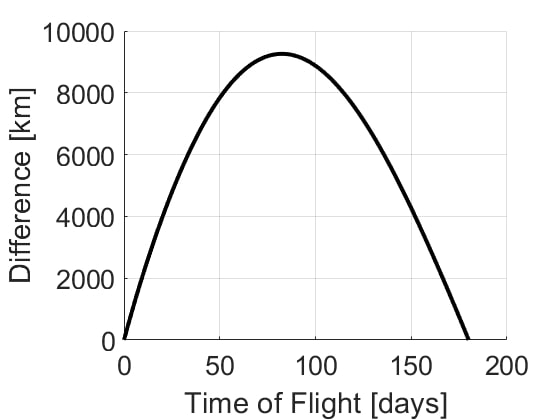}
        \caption{SRP vs unperturbed Lambert arcs.}
        \label{fig: SRPPertUnpertError}
    \end{subfigure}
    \begin{subfigure}{0.45\textwidth}
        \includegraphics[width=1\textwidth]{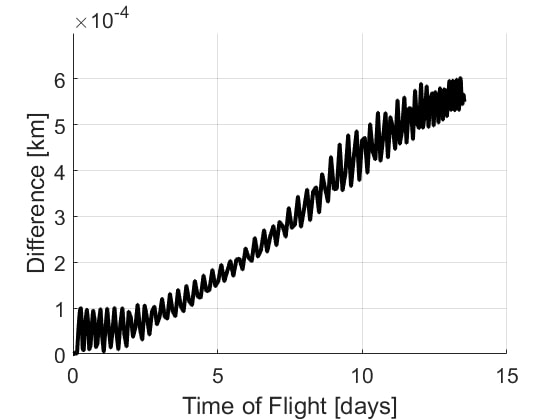}
        \caption{TFC vs DC algorithms.}
        \label{fig: SRPPertDiffCorrError}
    \end{subfigure}
    \caption{Error plots showing the norm of the vector difference between each state.}
    \label{fig: SRPerror}
\end{figure}

The final perturbation introduced in this investigation is the effects of SRP. Although the Sun's photons are massless, they are able to impart momentum to surfaces that they come into contact with. The magnitude and direction of the change in momentum due to SRP is given by Eq.~\eqref{eq:acc_srp}. The surface area of the chosen spacecraft is $20,000$ m$^2$, the reflectivity coefficient $\rho_s$ is $0.9$, the absorption $\rho_a$ and diffraction $\rho_d$ coefficients are $0.1$, the mass is $5$ kg, and the surface area normal direction is held constant in the $x$-axis direction. The parameters are selected to exaggerate the effects of the SRP. The same Earth-Venus transfer from Section~\ref{sec: unpert} is used in this test, using the same Legendre polynomial degree. The effects of SRP on this spacecraft, although minor, are clear from Figure~\ref{fig: SRPOrbit}. Figure~\ref{fig: SRPPertUnpertError} shows the magnitude of the effects, while Figure~\ref{fig: SRPPertDiffCorrError} validates the results using a DC algorithm.

\subsection{Perturbed Problem Performance Analysis}

\begin{figure}[t!]
\centering
\begin{subfigure}{.33\textwidth}
  \centering
  \includegraphics[width=1\linewidth]{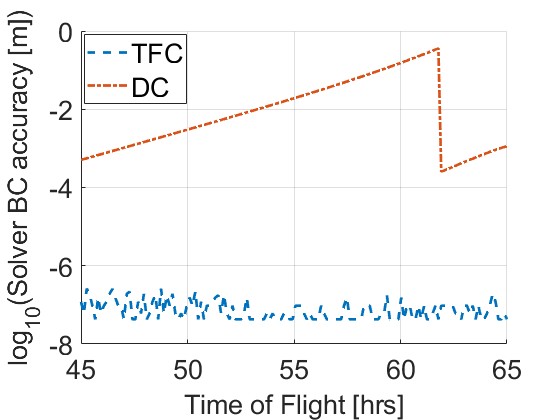}
  \caption{\label{TOF1pert}Final BC error vs ToF}
\end{subfigure}%
\begin{subfigure}{.33\textwidth}
  \centering
  \includegraphics[width=1\linewidth]{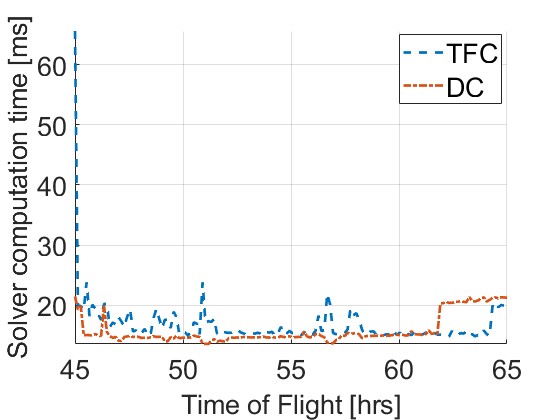}
  \caption{\label{TOF2pert}Computation time vs ToF}
\end{subfigure}
\begin{subfigure}{.33\textwidth}
  \centering
  \includegraphics[width=1\linewidth]{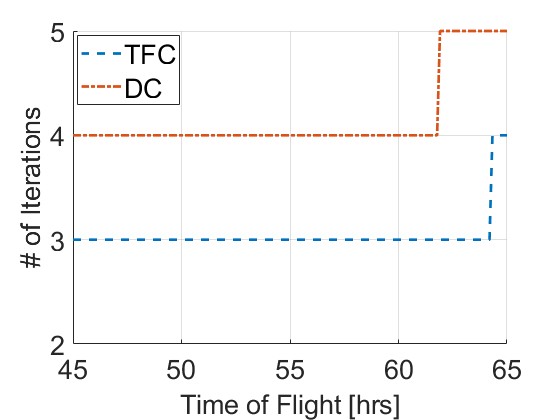}
  \caption{\label{TOF3pert}Iterations vs ToF}
\end{subfigure}
\begin{subfigure}{.33\textwidth}
  \centering
  \includegraphics[width=1\linewidth]{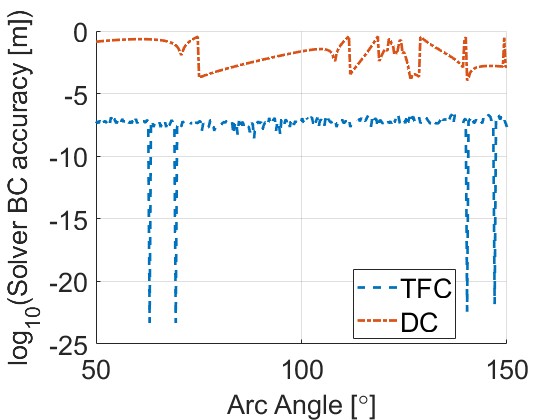}
  \caption{\label{ANG1pert}Final BC error vs $\vartheta_r$}
\end{subfigure}%
\begin{subfigure}{.33\textwidth}
  \centering
  \includegraphics[width=1\linewidth]{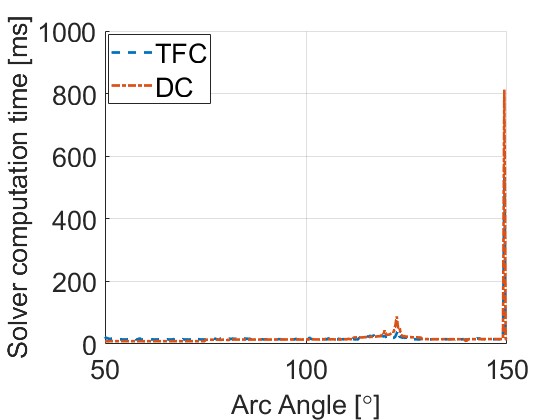}
  \caption{\label{ANG2pert}Computation time vs $\vartheta_r$}
\end{subfigure}
\begin{subfigure}{.33\textwidth}
  \centering
  \includegraphics[width=1\linewidth]{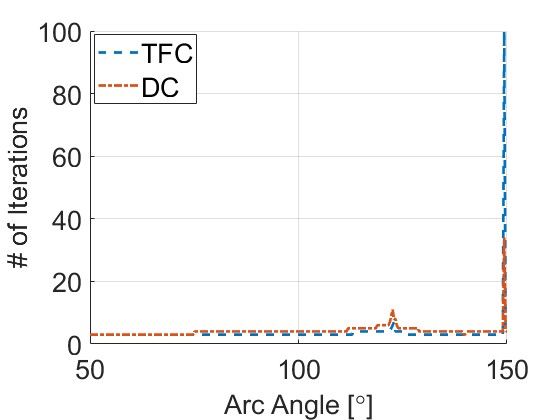}
  \caption{\label{ANG3pert}Iterations vs $\vartheta_r$}
\end{subfigure}
\begin{subfigure}{.33\textwidth}
  \centering
  \includegraphics[width=1\linewidth]{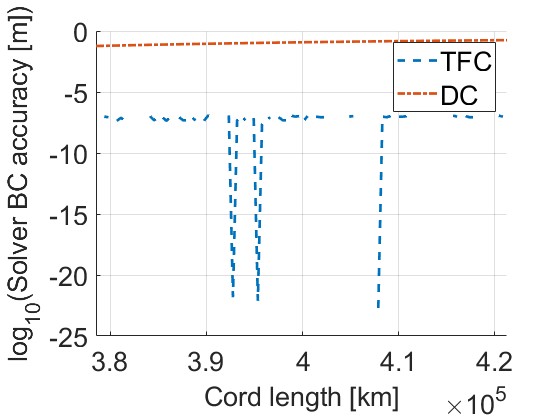}
  \caption{\label{CORD1pert}Final BC error vs cord}
\end{subfigure}%
\begin{subfigure}{.33\textwidth}
  \centering
  \includegraphics[width=1\linewidth]{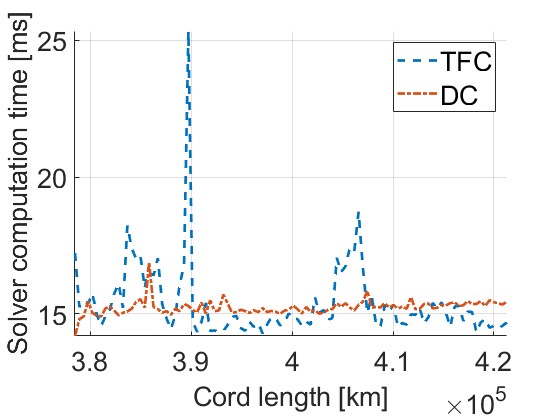}
  \caption{\label{CORD2pert}Computation time vs cord}
\end{subfigure}
\begin{subfigure}{.33\textwidth}
  \centering
  \includegraphics[width=1\linewidth]{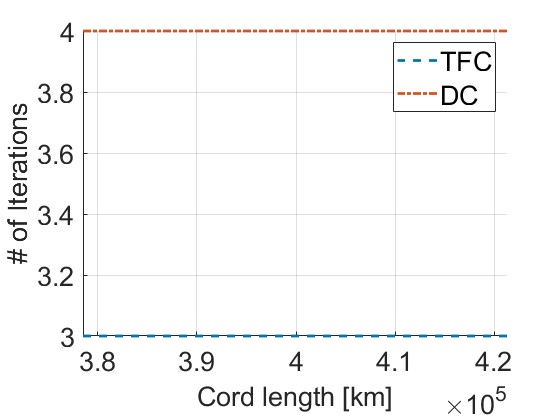}
  \caption{\label{CORD3pert}Iterations vs cord}
\end{subfigure}
\caption{\label{fig: pertPerformanceAnalysis}Performance analysis of perturbed TFC and DC for different Lambert arcs.}
\end{figure}

A similar procedure to Figure~\ref{fig:performanceAnalysis} is now performed to assess TFC's performance relative to DC (Figure~\ref{fig: pertPerformanceAnalysis}). In this case, the cislunar transfer from Figure~\ref{fig: 3BPOrbit} is used due to the presence of strong perturbations over an extended period of time. In contrast, the $J_2$ perturbation has a small effect over the comparatively short ToF, and the SRP perturbation has a smaller effect over a much longer ToF.
Figures~\ref{TOF1pert}-\ref{TOF3pert} analyze the effects of varying ToF between $45$ and $65$ hours while fixing the two BC with a transfer arc angle of $40^\circ$. Figures~\ref{ANG1pert}-\ref{ANG3pert} vary the transfer angle between $50^\circ$ and 150$^\circ$ while fixing the ToF to $60$ hours. Lastly, Figures~\ref{CORD1pert}-\ref{CORD3pert} fix the ToF to $60$ hours and the transfer angle to $110^\circ$, while varying the cord length from $362,000$ km (Moon's perigee) to $405,000$ km (Moon's apogee). All plots leads to the same conclusion: TFC is able to accurately and rapidly obtain a solution for the shown variable conditions. Figure~\ref{ANG3pert} shows a large spike in number of iterations due to the algorithm's singularity near 180$^\circ$. Figure~\ref{CORD2pert}, on the other hand, shows several spikes that seem to indicate conditions where TFC is slower than DC. These results, alongside Figures~\ref{CORD2} and ~\ref{CORD3}, leads to the investigation performed in the next section. The erratic behavior observed is a result of the polynomial type and degree not being suitable for the BC and ToF provided. Nevertheless, the main advantage of TFC with respect to DC is the lack of pre-computed initial guess. This provides a key advantage in solving perturbed problems which other solvers do not possess.

\section{Polynomial Analysis}\label{sec: poly}

Certain scenarios tested in this investigation appear to converge faster or slower due to a variety of reasons. Primarily, the magnitude of the initial and final positions appears to play a significant role in convergence. To solve this issue, most solutions are generated using non-dimensional units, where characteristic quantities vary depending on the context of the test. The reason for the divergence lies in the degree and type of the polynomial chosen. This reason led to an investigation into the effects of various parameters on the polynomial degree as well as type.

The Gegenbauer polynomial set is used to change polynomial type in a consistent manner. This set uses a selector variable $\alpha$ to control the polynomial type independently of the degree in the following recurrence relation:
\begin{align}
    C_0^{(\alpha)}(x) &= 1 \qquad
    C_1^{(\alpha)}(x) = 2\alpha x \\
    (n+1)C_{n+1}^{(\alpha)}(x) &=2(n+\alpha)xC_n^{(\alpha)}(x) - (n+2\alpha-1)C_{n-1}^{(\alpha)}(x) 
\end{align}
The Legendre polynomial corresponds to the Gegenbauer set with $\alpha = 0.5$; $\alpha$ may be incremented by a multiple of $0.5$ to modify the polynomial type (minimum of $\alpha =-0.5$). The subscript of each constant $C$ represents the degree of the polynomial; for a second-degree iteration, $n=1$.

\begin{figure}[h!]
    \centering
    \includegraphics[width=0.9\textwidth]{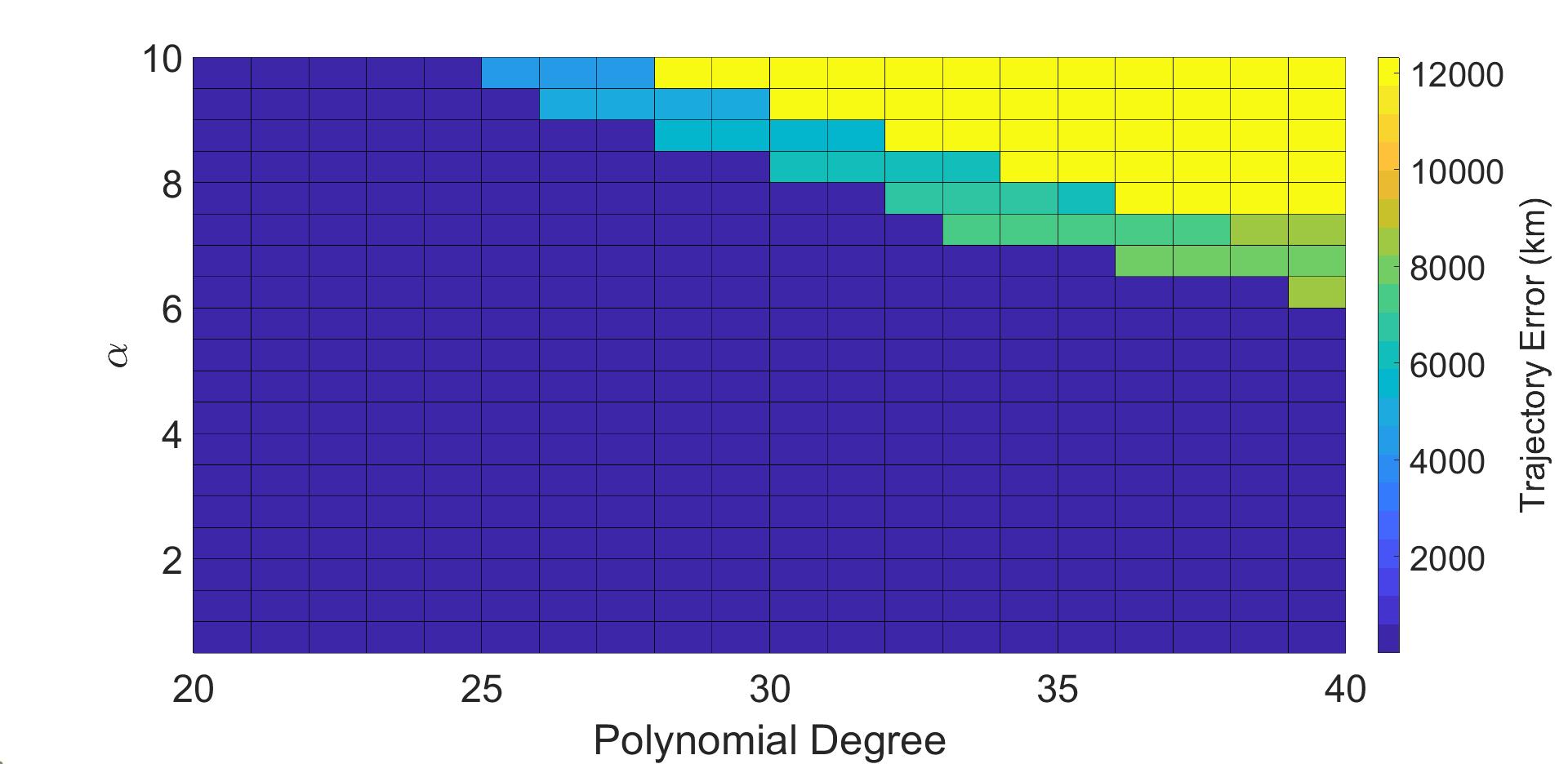}
    \caption{Degree-$\alpha$ plot at $r_f/r_0 = 1$ (8378 km) showing the expected trend.}
    \label{fig: deg-alpha-norm}
\end{figure}

\begin{figure}[h!]
    \centering
    \includegraphics[width=0.9\textwidth]{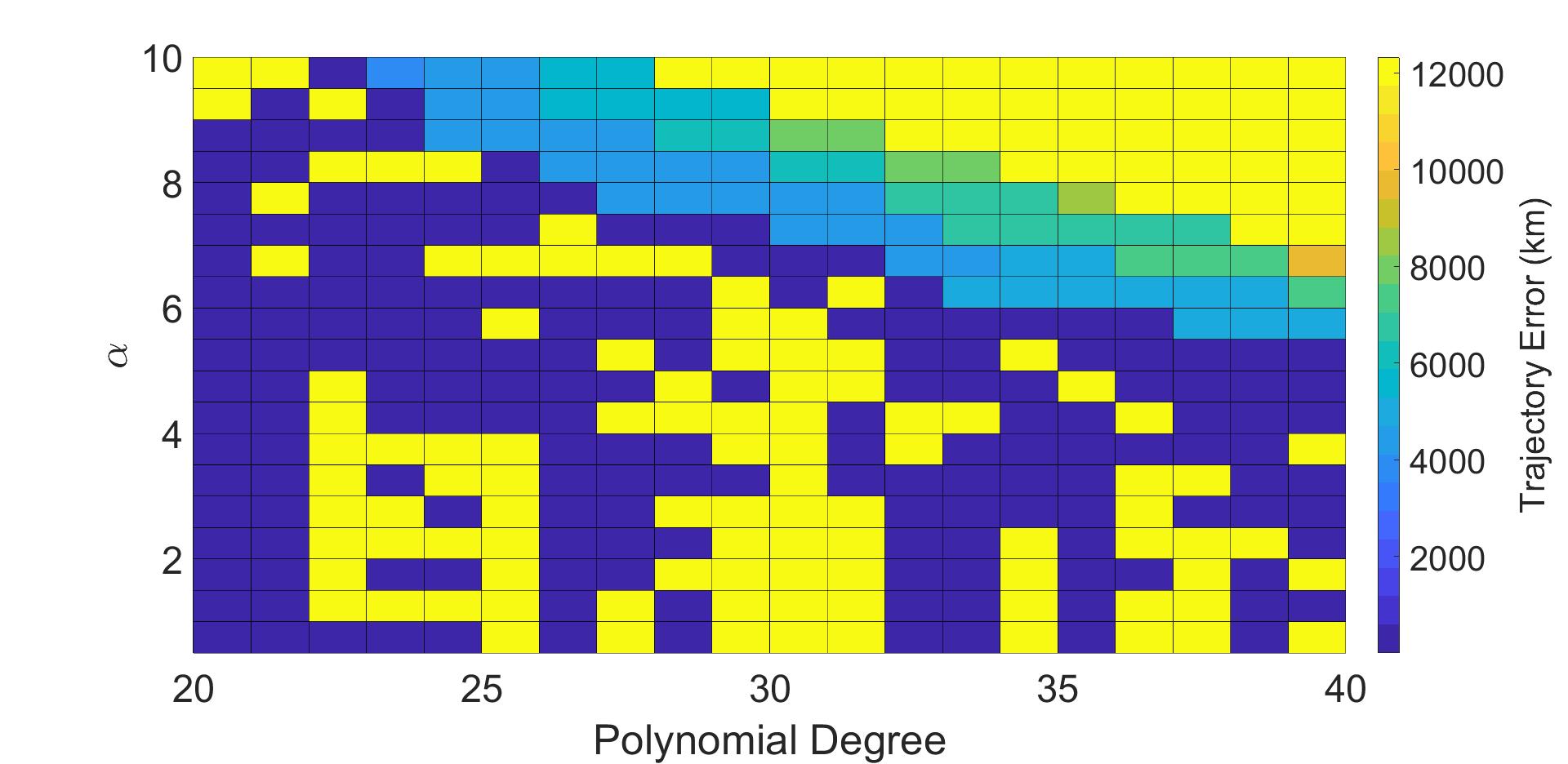}
    \caption{Degree-$\alpha$ plot at $r_f/r_0 = 2.32$ (19410 km) showing failures to converge.}
    \label{fig: deg-alpha-chaos}
\end{figure}

\begin{figure}[h!]
    \centering
    \includegraphics[width=0.9\textwidth]{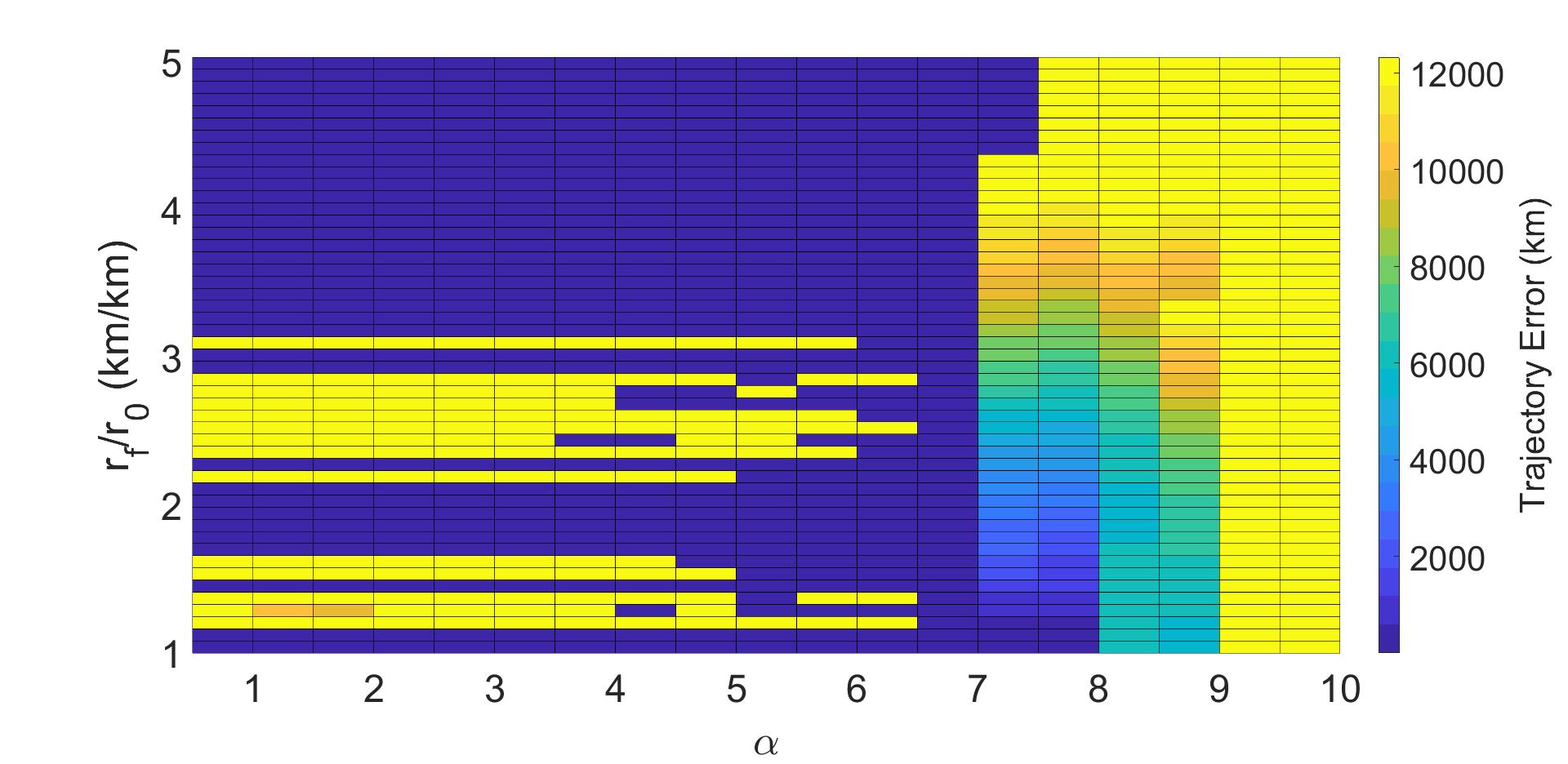}
    \caption{$\alpha$-radius plot at degree 30 showing sections of concentrated errors.}
    \label{fig: alpha-z}
\end{figure}

To analyze the effects of varying the polynomial type and degree on convergence, Figures~\ref{CORD2} and~\ref{CORD3} are analyzed in depth. Recall that the angle and ToF are fixed at $120^\circ$ and $2.5$ hours, respectively, while the trajectory is varied by changing the final radius from $8,378$ km to $42,164$ km. The following test is conducted for multiple combinations of degree and $\alpha$; the degree ranges from $20$ to $40$, and $\alpha$ ranges from $0.5$ to $10$. (Figures~\ref{fig: deg-alpha-norm}, ~\ref{fig: deg-alpha-chaos}, ~\ref{fig: alpha-z}). Note that the color bar measures the error of the trajectory. This error is calculated by propagating the output initial velocity from TFC with a two-body propagator, and then taking the distance between the input final point and the propagated destination. Figures ~\ref{fig: deg-alpha-norm} and ~\ref{fig: deg-alpha-chaos} show the expected behavior and errors, respectively: most degree-$\alpha$ pairings would result in relatively low error, with the error increasing as the polynomial becomes overly complex; note that the convergence algorithm is limited to $200$ iterations in this test. However, at certain settings, many pairings result in failures to converge, leading to extremely high error. As Figure~\ref{fig: alpha-z} shows, this behavior is erratic and unpredictable, but limited to specific values of $r_f/r_0$, i.e., horizontal segments. Additionally, certain degree-$\alpha$ pairings result in reduced quantities of errors, potentially avoiding them entirely. 
Generally, higher error trajectories are correlated with iteration number, as they struggle to converge within the given tolerance. When the iteration limit has been reached, TFC is not able to find a solution with the given polynomial type, degree, tolerance, or ToF. Due to this, it is fair to say that TFC is in some cases limited by this drawback. It is important to note that this error appears in very limited cases dependent on the polynomial chosen as shown in Figures~\ref{fig: deg-alpha-norm} - \ref{fig: alpha-z}. Thus, knowledge of the relationship between error and polynomial degree would reduce the risk of divergence. At this time, no mathematical description relating error and degree is deduced, although a deeper investigation into this topic is possible.

\section{Concluding Remarks}

The work presented in this investigation highlights a novel method for solving Lambert's problem using the Theory of Functional Connections (TFC). Unperturbed cases are presented and validated using a differential corrections (DC) algorithm and a robust Lambert solver (RLS) that leverages Izzo's as well as Blanchard's algorithms. Comparing the three algorithms shows that RLS is substantially faster than the two numerical methods, but TFC guarantees the final BC to a very small error. Perturbed cases are also shown and compared with DC. Results show that, although DC is faster in cases where a good initial guess is given or the angle between boundaries is close to 180$^\circ$, TFC is generally faster. Notably, no initial guess is necessary for TFC, while DC requires a good guess for it to be as efficient as TFC. It should be noted that DC always requires an initial guess that directly impacts computation time, while TFC's initial guess is typically straightforward. Multi-revolution orbits are also presented in perturbed cases. Throughout this investigation, 200 discretization points are used to generate solutions, without causing any issues for computation time or accuracy. Finally, an in-depth polynomial analysis shows the effects of the type and degree used on convergence time for various scenarios.

\section{Acknowledgment}

The research was partially supported by the Embry-Riddle Aeronautical University Faculty Research Development Program and GAANN fellowship. Assistance from colleagues in the Space Trajectories and Applications Research group at Embry-Riddle Aeronautical University is acknowledged as is the support from the Aerospace Engineering department.

\clearpage
\bibliography{Preprint}


\begin{thebibliography}{30}
\ifx \bisbn   \undefined \def \bisbn  #1{ISBN #1}\fi
\ifx \binits  \undefined \def \binits#1{#1}\fi
\ifx \bauthor  \undefined \def \bauthor#1{#1}\fi
\ifx \batitle  \undefined \def \batitle#1{#1}\fi
\ifx \bjtitle  \undefined \def \bjtitle#1{#1}\fi
\ifx \bvolume  \undefined \def \bvolume#1{\textbf{#1}}\fi
\ifx \byear  \undefined \def \byear#1{#1}\fi
\ifx \bissue  \undefined \def \bissue#1{#1}\fi
\ifx \bfpage  \undefined \def \bfpage#1{#1}\fi
\ifx \blpage  \undefined \def \blpage #1{#1}\fi
\ifx \burl  \undefined \def \burl#1{\textsf{#1}}\fi
\ifx \doiurl  \undefined \def \doiurl#1{\url{https://doi.org/#1}}\fi
\ifx \betal  \undefined \def \betal{\textit{et al.}}\fi
\ifx \binstitute  \undefined \def \binstitute#1{#1}\fi
\ifx \binstitutionaled  \undefined \def \binstitutionaled#1{#1}\fi
\ifx \bctitle  \undefined \def \bctitle#1{#1}\fi
\ifx \beditor  \undefined \def \beditor#1{#1}\fi
\ifx \bpublisher  \undefined \def \bpublisher#1{#1}\fi
\ifx \bbtitle  \undefined \def \bbtitle#1{#1}\fi
\ifx \bedition  \undefined \def \bedition#1{#1}\fi
\ifx \bseriesno  \undefined \def \bseriesno#1{#1}\fi
\ifx \blocation  \undefined \def \blocation#1{#1}\fi
\ifx \bsertitle  \undefined \def \bsertitle#1{#1}\fi
\ifx \bsnm \undefined \def \bsnm#1{#1}\fi
\ifx \bsuffix \undefined \def \bsuffix#1{#1}\fi
\ifx \bparticle \undefined \def \bparticle#1{#1}\fi
\ifx \barticle \undefined \def \barticle#1{#1}\fi
\bibcommenthead
\ifx \bconfdate \undefined \def \bconfdate #1{#1}\fi
\ifx \botherref \undefined \def \botherref #1{#1}\fi
\ifx \url \undefined \def \url#1{\textsf{#1}}\fi
\ifx \bchapter \undefined \def \bchapter#1{#1}\fi
\ifx \bbook \undefined \def \bbook#1{#1}\fi
\ifx \bcomment \undefined \def \bcomment#1{#1}\fi
\ifx \oauthor \undefined \def \oauthor#1{#1}\fi
\ifx \citeauthoryear \undefined \def \citeauthoryear#1{#1}\fi
\ifx \endbibitem  \undefined \def \endbibitem {}\fi
\ifx \bconflocation  \undefined \def \bconflocation#1{#1}\fi
\ifx \arxivurl  \undefined \def \arxivurl#1{\textsf{#1}}\fi
\csname PreBibitemsHook\endcsname

\bibitem[\protect\citeauthoryear{de~la Torre~Sangra and Fantino}{2019}]{Fantino2021}
\begin{botherref}
\oauthor{\bsnm{Torre~Sangra}, \binits{D.}},
\oauthor{\bsnm{Fantino}, \binits{E.}}:
{Review of Lambert's Problem}.
25th International Symposium on Space Flight Dynamics
(2019)
\end{botherref}
\endbibitem

\bibitem[\protect\citeauthoryear{Klumpp}{1990}]{Lumpp1990}
\begin{botherref}
\oauthor{\bsnm{Klumpp}, \binits{A.R.}}:
{Universal Lambert and Kepler Algorithms for Autonomous Rendezvous}.
Astrodynamics Conference
(1990)
\end{botherref}
\endbibitem

\bibitem[\protect\citeauthoryear{Peterson et~al.}{2010}]{Peterson2010}
\begin{barticle}
\bauthor{\bsnm{Peterson}, \binits{G.E.}},
\bauthor{\bsnm{Campbell}, \binits{E.T.}},
\bauthor{\bsnm{Hawkins}, \binits{A.M.}},
\bauthor{\bsnm{Balbás}, \binits{J.}},
\bauthor{\bsnm{Ivy}, \binits{S.}},
\bauthor{\bsnm{Merkurjev}, \binits{E.}},
\bauthor{\bsnm{Hall}, \binits{T.}},
\bauthor{\bsnm{Rodriguez}, \binits{P.}},
\bauthor{\bsnm{Gustafson}, \binits{E.D.}}:
\batitle{{Relative Performance of Lambert Solvers 1: Zero-Revolution Methods}}.
\bjtitle{Advances in the Astronautical Sciences}
\bvolume{136},
\bfpage{1495}--\blpage{1509}
(\byear{2010})
\end{barticle}
\endbibitem

\bibitem[\protect\citeauthoryear{il~Lee et~al.}{2016}]{Lee2016}
\begin{barticle}
\bauthor{\bsnm{Lee}, \binits{S.-i.}},
\bauthor{\bsnm{Ahn}, \binits{J.}},
\bauthor{\bsnm{Bang}, \binits{J.}}:
\batitle{{Dynamic Selection of Zero-Revolution Lambert Algorithms Using Performance Comparison Map}}.
\bjtitle{Aerospace Science and Technology}
\bvolume{51},
\bfpage{96}--\blpage{105}
(\byear{2016})
\doiurl{10.1016/j.ast.2016.01.018}
\end{barticle}
\endbibitem

\bibitem[\protect\citeauthoryear{Wailliez}{2014}]{Wailliez2014}
\begin{barticle}
\bauthor{\bsnm{Wailliez}, \binits{S.E.}}:
\batitle{{On Lambert’s Problem and the Elliptic Time of Flight Equation: A Simple Semi-Analytical Inversion Method}}.
\bjtitle{Advances in Space Research}
\bvolume{53}(\bissue{5}),
\bfpage{890}--\blpage{898}
(\byear{2014})
\doiurl{10.1016/j.asr.2013.12.033}
\end{barticle}
\endbibitem

\bibitem[\protect\citeauthoryear{Thorne and Bain}{1995}]{Thorne1995}
\begin{barticle}
\bauthor{\bsnm{Thorne}, \binits{J.D.}},
\bauthor{\bsnm{Bain}, \binits{R.D.}}:
\batitle{{Series Reversion/Inversion of Lambert's Time Function}}.
\bjtitle{Journal of the Astronautical Sciences}
\bvolume{43}(\bissue{3}),
\bfpage{277}--\blpage{287}
(\byear{1995})
\doiurl{10.2514/6.1990-2886}
\end{barticle}
\endbibitem

\bibitem[\protect\citeauthoryear{Thorne}{2014}]{Thorne2014}
\begin{botherref}
\oauthor{\bsnm{Thorne}, \binits{J.D.}}:
{Convergence Behavior of Series Solutions of the Lambert Problem}.
Journal of Guidance, Control, and Dynamics
(8),
1--6
(2014)
\doiurl{10.2514/1.G000701}
\end{botherref}
\endbibitem

\bibitem[\protect\citeauthoryear{Avanzini}{2008}]{Avanzini2008}
\begin{barticle}
\bauthor{\bsnm{Avanzini}, \binits{G.}}:
\batitle{{A Simple Lambert Algorithm}}.
\bjtitle{Journal of Guidance Control and Dynamics}
\bvolume{31},
\bfpage{1587}--\blpage{1594}
(\byear{2008})
\doiurl{10.2514/1.36426}
\end{barticle}
\endbibitem

\bibitem[\protect\citeauthoryear{Boltz}{1984}]{Boltz1984}
\begin{barticle}
\bauthor{\bsnm{Boltz}, \binits{F.W.}}:
\batitle{{Second-Order P-Iterative Solution of the Lambert/Gauss Problem}}.
\bjtitle{Journal of the Astronautical Sciences}
\bvolume{32},
\bfpage{475}--\blpage{485}
(\byear{1984})
\end{barticle}
\endbibitem

\bibitem[\protect\citeauthoryear{Arlulkar and D.}{2011}]{Arlulkar2011}
\begin{barticle}
\bauthor{\bsnm{Arlulkar}, \binits{P.}},
\bauthor{\bsnm{D.}, \binits{N.}}:
\batitle{{Solution Based on Dynamical Approach for Multiple-Revolution Lambert Problem}}.
\bjtitle{Journal of Guidance Control and Dynamics}
\bvolume{34},
\bfpage{920}--\blpage{923}
(\byear{2011})
\doiurl{10.2514/1.51723}
\end{barticle}
\endbibitem

\bibitem[\protect\citeauthoryear{Curtis}{2020}]{curtisbook}
\begin{bbook}
\bauthor{\bsnm{Curtis}, \binits{H.D.}}:
\bbtitle{"Orbital Mechanics for Engineering Students"}.
\bpublisher{Springer},
\blocation{Amsterdam}
(\byear{2020})
\end{bbook}
\endbibitem

\bibitem[\protect\citeauthoryear{Vallado and McClain}{2013}]{vallado2001fundamentals}
\begin{bbook}
\bauthor{\bsnm{Vallado}, \binits{D.A.}},
\bauthor{\bsnm{McClain}, \binits{W.D.}}:
\bbtitle{"Fundamentals of Astrodynamics and Applications"}.
\bpublisher{Microcosm Press},
\blocation{Hawthorne}
(\byear{2013})
\end{bbook}
\endbibitem

\bibitem[\protect\citeauthoryear{Battin}{1999}]{battinbook}
\begin{bbook}
\bauthor{\bsnm{Battin}, \binits{R.H.}}:
\bbtitle{"An Introduction to the Mathematics and Methods of Astrodynamics"}.
\bpublisher{American Institute of Aeronautics and Astronautics},
\blocation{Reston}
(\byear{1999})
\end{bbook}
\endbibitem

\bibitem[\protect\citeauthoryear{Izzo}{2014}]{Izzo2014}
\begin{barticle}
\bauthor{\bsnm{Izzo}, \binits{D.}}:
\batitle{{Revisiting Lambert's Problem}}.
\bjtitle{Celestial Mechanics and Dynamical Astronomy}
\bvolume{120}(\bissue{1}),
\bfpage{1}--\blpage{15}
(\byear{2014})
\doiurl{10.1007/s10569-014-9587-y}
\end{barticle}
\endbibitem

\bibitem[\protect\citeauthoryear{{Gooding}}{1990}]{Gooding1990}
\begin{barticle}
\bauthor{\bsnm{{Gooding}}, \binits{R.H.}}:
\batitle{{A Procedure for the Solution of Lambert's Orbital Boundary-Value Problem}}.
\bjtitle{Celestial Mechanics and Dynamical Astronomy}
\bvolume{48}(\bissue{2}),
\bfpage{145}--\blpage{165}
(\byear{1990})
\doiurl{10.1007/BF00049511}
\end{barticle}
\endbibitem

\bibitem[\protect\citeauthoryear{Blanchard and Lancaster}{1968}]{Lancaster1968}
\begin{bchapter}
\bauthor{\bsnm{Blanchard}, \binits{R.}},
\bauthor{\bsnm{Lancaster}, \binits{E.R.}}:
\bctitle{{A Unified Form of Lambert's Theorem}}.
In: \bbtitle{NASA Technical Note TN D-5368}
(\byear{1968})
\end{bchapter}
\endbibitem

\bibitem[\protect\citeauthoryear{Russell}{2021}]{Russell2021}
\begin{barticle}
\bauthor{\bsnm{Russell}, \binits{R.}}:
\batitle{{Complete Lambert Solver Including Second-Order Sensitivities}}.
\bjtitle{Acta Astronautica}
\bvolume{178},
\bfpage{312}--\blpage{321}
(\byear{2021})
\doiurl{10.2514/1.G006089}
\end{barticle}
\endbibitem

\bibitem[\protect\citeauthoryear{{De La Torre} et~al.}{2018}]{Fantino2018}
\begin{barticle}
\bauthor{\bsnm{{De La Torre}}, \binits{D.}},
\bauthor{\bsnm{Flores}, \binits{R.}},
\bauthor{\bsnm{Fantino}, \binits{E.}}:
\batitle{{On the Solution of Lambert's Problem by Regularization}}.
\bjtitle{Acta Astronautica}
\bvolume{153},
\bfpage{26}--\blpage{38}
(\byear{2018})
\doiurl{10.1016/j.actaastro.2018.10.010}
\end{barticle}
\endbibitem

\bibitem[\protect\citeauthoryear{Simo}{1973}]{Simo1973}
\begin{barticle}
\bauthor{\bsnm{Simo}, \binits{C.}}:
\batitle{{Solucion al Problema de Lambert Mediante Regularizacion}}.
\bjtitle{Collectanea Mathematica}
\bvolume{24}(\bissue{3}),
\bfpage{231}--\blpage{247}
(\byear{1973})
\end{barticle}
\endbibitem

\bibitem[\protect\citeauthoryear{Canales}{2021}]{Canales2021}
\begin{botherref}
\oauthor{\bsnm{Canales}, \binits{D.}}:
{Transfer Design Methodology Between Neighborhoods of Planetary Moons in the Circular Restricted Three-body Problem}.
PhD thesis,
Purdue University
(2021)
\end{botherref}
\endbibitem

\bibitem[\protect\citeauthoryear{Panicucci et~al.}{2018}]{Panicucci2018}
\begin{bchapter}
\bauthor{\bsnm{Panicucci}, \binits{P.}},
\bauthor{\bsnm{Morand}, \binits{V.}},
\bauthor{\bsnm{Hautesserres}, \binits{D.}}:
\bctitle{{Perturbed Lambert’s Problem Solver based on Differential Algebra Optimization}}.
In: \bbtitle{69th International Astronautical Congress}
(\byear{2018})
\end{bchapter}
\endbibitem

\bibitem[\protect\citeauthoryear{Woollands et~al.}{2015}]{Woollands2015}
\begin{barticle}
\bauthor{\bsnm{Woollands}, \binits{R.M.}},
\bauthor{\bsnm{Bani~Younes}, \binits{A.}},
\bauthor{\bsnm{Junkins}, \binits{J.L.}}:
\batitle{{New Solutions for the Perturbed Lambert Problem Using Regularization and Picard Iteration}}.
\bjtitle{Journal of Guidance, Control, and Dynamics}
\bvolume{38}(\bissue{9}),
\bfpage{1548}--\blpage{1562}
(\byear{2015})
\doiurl{10.2514/1.G001028}
\end{barticle}
\endbibitem

\bibitem[\protect\citeauthoryear{Yang et~al.}{2022}]{Yang2022}
\begin{barticle}
\bauthor{\bsnm{Yang}, \binits{B.}},
\bauthor{\bsnm{Li}, \binits{S.}},
\bauthor{\bsnm{Feng}, \binits{J.}},
\bauthor{\bsnm{Vasile}, \binits{M.}}:
\batitle{{Fast Solver for J2-Perturbed Lambert Problem Using Deep Neural Network}}.
\bjtitle{Journal of Guidance, Control, and Dynamics}
\bvolume{45}(\bissue{5}),
\bfpage{875}--\blpage{884}
(\byear{2022})
\doiurl{10.2514/1.G006091}
\end{barticle}
\endbibitem

\bibitem[\protect\citeauthoryear{Leake et~al.}{2022}]{TFC_Book}
\begin{bbook}
\bauthor{\bsnm{Leake}, \binits{C.}},
\bauthor{\bsnm{Johnston}, \binits{H.}},
\bauthor{\bsnm{Mortari}, \binits{D.}}:
\bbtitle{"The Theory of Functional Connections: A Functional Interpolation. Framework with Applications"}.
\bpublisher{Lulu},
\blocation{Morrisville NC}
(\byear{2022})
\end{bbook}
\endbibitem

\bibitem[\protect\citeauthoryear{Mortari et~al.}{2023}]{Criscola2023}
\begin{bchapter}
\bauthor{\bsnm{Mortari}, \binits{D.}},
\bauthor{\bsnm{Criscola}, \binits{F.}},
\bauthor{\bsnm{Canales}, \binits{D.}}:
\bctitle{{(Preprint) Perturbed Lambert Problem using the Theory of Functional Connections}}.
In: \bbtitle{33rd AAS/AIAA Space Flight Mechanics Meeting, Austin, Texas}
(\byear{2023})
\end{bchapter}
\endbibitem

\bibitem[\protect\citeauthoryear{Criscola et~al.}{2023}]{Criscola2023_2}
\begin{bchapter}
\bauthor{\bsnm{Criscola}, \binits{F.}},
\bauthor{\bsnm{Hudson}, \binits{Z.}},
\bauthor{\bsnm{Canales}, \binits{D.}},
\bauthor{\bsnm{Mortari}, \binits{D.}}:
\bctitle{{(Preprint) Solution of the Perturbed Lambert's Problem using the Theory of Functional Connections}}.
In: \bbtitle{AAS/AIAA Astrodynamics Specialist Meeting, Big Sky, Montana}
(\byear{2023})
\end{bchapter}
\endbibitem

\bibitem[\protect\citeauthoryear{Johnston and Mortari}{2018}]{Johnston2018}
\begin{bchapter}
\bauthor{\bsnm{Johnston}, \binits{H.}},
\bauthor{\bsnm{Mortari}, \binits{D.}}:
\bctitle{{(Preprint) The Theory of Connections Applied to Perturbed Lambert's Problem}}.
In: \bbtitle{AAS/AIAA Astrodynamics Specialist Meeting, Snowbird, Utah}
(\byear{2018})
\end{bchapter}
\endbibitem

\bibitem[\protect\citeauthoryear{Mortari}{2017a}]{U-TFC}
\begin{barticle}
\bauthor{\bsnm{Mortari}, \binits{D.}}:
\batitle{{The Theory of Connections: Connecting Points}}.
\bjtitle{Mathematics}
\bvolume{5}(\bissue{4}),
\bfpage{1}--\blpage{15}
(\byear{2017})
\doiurl{10.3390/math5040057}
\end{barticle}
\endbibitem

\bibitem[\protect\citeauthoryear{Mortari}{2017b}]{LDE}
\begin{barticle}
\bauthor{\bsnm{Mortari}, \binits{D.}}:
\batitle{{Least-Squares Solution of Linear Differential Equations}}.
\bjtitle{Mathematics}
\bvolume{5}(\bissue{4}),
\bfpage{1}--\blpage{18}
(\byear{2017})
\doiurl{10.3390/math5040048}
\end{barticle}
\endbibitem

\bibitem[\protect\citeauthoryear{Zardain et~al.}{2020}]{zardain}
\begin{bchapter}
\bauthor{\bsnm{Zardain}, \binits{L.}},
\bauthor{\bsnm{Farr{\'e}s}, \binits{A.}},
\bauthor{\bsnm{Puig}, \binits{A.}}:
\bctitle{{High-fidelity Modeling and Visualizing of Solar Radiation Pressure: a Framework for High-Fidelity Analysis}}.
In: \bbtitle{2020 AAS/AIAA Astrodynamics Specialist Conference}
(\byear{2020})
\end{bchapter}
\endbibitem

\end{thebibliography}

\newpage
\appendix

\appendix
\section{Partial Derivatives}
\label{sec: appendixPartials}

Starting with the partial derivatives of the functionals with respect to $\B{\xi}_p$, $\B{\xi}_{\vartheta}$, and $\B{\xi}_h$:
\begin{align}
    \dfrac{\partial p}{\partial \B{\xi}_p} = \dfrac{\partial \vartheta}{\partial \B{\xi}_{\vartheta}} = \dfrac{\partial h}{\partial \B{\xi}_h} &= \B{s} - \dfrac{\Delta T - t}{\Delta T} \, \B{s_0} - \dfrac{t}{\Delta T} \, \B{s_f} \\
    \dfrac{\partial \dot{p}}{\partial \B{\xi}_p} = \dfrac{\partial \dot{\vartheta}}{\partial \B{\xi}_{\vartheta}} = \dfrac{\partial \dot{h}}{\partial \B{\xi}_h} &= \dot{\B{s}} + \dfrac{\B{s_0}}{\Delta T} - \dfrac{\B{s_f}}{\Delta T} \\
    \dfrac{\partial \ddot{p}}{\partial \B{\xi}_p} = \dfrac{\partial \ddot{\vartheta}}{\partial \B{\xi}_{\vartheta}} = \dfrac{\partial \ddot{h}}{\partial \B{\xi}_h} &= \ddot{\B{s}}
\end{align}
which are used in the derivatives of $r$ and $\B{r}$:
\begin{align}
    \dfrac{\partial r}{\partial \B{\xi}_p} &= \dfrac{p}{\sqrt{p^2 + h^2}} \, \dfrac{\partial p}{\partial \B{\xi}_p} \\
    \dfrac{\partial r}{\partial \B{\xi}_h} &= \dfrac{h}{\sqrt{p^2 + h^2}} \, \dfrac{\partial h}{\partial \B{\xi}_h} \\
    \dfrac{\partial \B{r}}{\partial \B{\xi}_p} &= \dfrac{\partial p}{\partial \B{\xi}_p} \, \big(\hat{\bm r}_0 \cos\vartheta + \hat{\bm t}_0 \sin\vartheta\big) \\
    \dfrac{\partial \B{r}}{\partial \B{\xi}_{\vartheta}} &= p \dfrac{\partial \vartheta}{\partial \B{\xi}_{\vartheta}} \big(-\hat{\bm r}_0 \sin\vartheta + \hat{\bm t}_0 \cos \vartheta\big) \\
    \dfrac{\partial \B{r}}{\partial \B{\xi}_h} &= \dfrac{\partial h}{\partial \B{\xi}_h} \hat{\bm h}_0
\end{align}
The partials of $\dot{\B{r}}$ and $\ddot{\B{r}}$ are
\begin{align}
   \dfrac{\partial \dot{\B{r}}}{\partial \B{\xi}_p} &= \bigg[\dfrac{\partial \dot{p}}{\partial \B{\xi}_p} \cos\vartheta - \dfrac{\partial p}{\partial \B{\xi}_p} \dot{\vartheta} \sin\vartheta\bigg] \hat{\bm r}_0 + \bigg[\dfrac{\partial \dot{p}}{\partial \B{\xi}_p} \sin\vartheta + \dfrac{\partial p}{\partial \B{\xi}_p} \dot{\vartheta} \cos\vartheta\bigg] \hat{\bm t}_0 \\
   \dfrac{\partial \dot{\B{r}}}{\partial \B{\xi}_{\vartheta}} &= \bigg[-\dot{p} \dfrac{\partial \vartheta}{\partial \B{\xi}_{\vartheta}} \sin\vartheta - p \dfrac{\partial \dot{\vartheta}}{\partial \B{\xi}_{\vartheta}} \sin\vartheta - p \dot{\vartheta} \dfrac{\partial \vartheta}{\partial \B{\xi}_{\vartheta}} \cos\vartheta\bigg] \hat{\bm r}_0 + \\
   & \, +\bigg[\dot{p} \dfrac{\partial \vartheta}{\partial \B{\xi}_{\vartheta}} \cos\vartheta + p \dfrac{\partial \dot{\vartheta}}{\partial \B{\xi}_{\vartheta}} \cos\vartheta - p \dfrac{\partial \vartheta}{\partial \B{\xi}_{\vartheta}} \dot{\vartheta} \sin\vartheta\bigg] \hat{\bm t}_0 \nonumber \\
   \dfrac{\partial \dot{\B{r}}}{\partial \B{\xi}_h} &= \dfrac{\partial \dot{h}}{\partial \B{\xi}_h} \hat{\bm h}_0
\end{align}
and
\begin{align}
    \dfrac{\partial \ddot{\B{r}}}{\partial \B{\xi}_p} &= \left[\left(\dfrac{\partial \ddot{p}}{\partial \B{\xi}_p} - \dfrac{\partial p}{\partial \B{\xi}_p} \dot{\vartheta}^2\right) \cos\vartheta - \left(2\dfrac{\partial \dot{p}}{\partial \B{\xi}_p} \dot{\vartheta} + \dfrac{\partial p}{\partial \B{\xi}_p} \ddot{\vartheta}\right) \sin\vartheta\right] \hat{\bm r}_0 + \\
    ~ &~ \;\; + \left[\left(\dfrac{\partial \ddot{p}}{\partial \B{\xi}_p} - \dfrac{\partial p}{\partial \B{\xi}_p} \dot{\vartheta}^2\right) \sin\vartheta + \left(2\dfrac{\partial \dot{p}}{\partial \B{\xi}_p} \dot{\vartheta} + \dfrac{\partial p}{\partial \B{\xi}_p} \ddot{\vartheta}\right) \cos\vartheta\right] \hat{\bm t}_0 \nonumber \\
    \dfrac{\partial \ddot{\B{r}}}{\partial \B{\xi}_{\vartheta}} &= \left[- 2 p \dot{\vartheta} \dfrac{\partial \dot{\vartheta}}{\partial \B{\xi}_{\vartheta}} \cos\vartheta - \left(2\dot{p} \dfrac{\partial \dot{\vartheta}}{\partial \B{\xi}_{\vartheta}} + p \dfrac{\partial \ddot{\vartheta}}{\partial \B{\xi}_{\vartheta}}\right) \sin\vartheta\right] \hat{\bm r}_0 + \\
    ~ &~ \;\; + \left[- 2 p \dot{\vartheta} \dfrac{\partial \dot{\vartheta}}{\partial \B{\xi}_{\vartheta}} \sin\vartheta + \left(2\dot{p} \dfrac{\partial \dot{\vartheta}}{\partial \B{\xi}_{\vartheta}} + p \dfrac{\partial \ddot{\vartheta}}{\partial \B{\xi}_{\vartheta}}\right) \cos\vartheta\right] \hat{\bm t}_0 \nonumber \\
    ~ &~ \;\; - \dfrac{\partial \vartheta}{\partial \B{\xi}_{\vartheta}} \left[\left(\ddot{p} - p \dot{\vartheta}^2\right) \sin\vartheta + \left(2\dot{p} \dot{\vartheta} + p \ddot{\vartheta}\right) \cos\vartheta\right] \hat{\bm r}_0 + \nonumber \\
    ~ &~ \;\; + \dfrac{\partial \vartheta}{\partial \B{\xi}_{\vartheta}} \left[\left(\ddot{p} - p \dot{\vartheta}^2\right) \cos\vartheta - \left(2\dot{p} \dot{\vartheta} + p \ddot{\vartheta}\right) \sin\vartheta\right] \hat{\bm t}_0 \nonumber \\
    \dfrac{\partial \ddot{\B{r}}}{\partial \B{\xi}_h} &= \dfrac{\partial \ddot{h}}{\partial \B{\xi}_h} \hat{\bm h}_0 
\end{align}
The Jacobian is finally computed with the above formulas:
\begin{subequations}
    \begin{alignat}{2}
    \label{eq:loss_xir}
    \dfrac{\partial \mathcal{L}}{\partial \B{\xi}_p} &= \dfrac{\partial \ddot{\B{r}}}{\partial \B{\xi}_p} - 3 \mu \dfrac{\B{r}}{r^4}\dfrac{\partial r}{\partial \B{\xi}_p} + \dfrac{\mu}{r^3}\dfrac{\partial \B{r}}{\partial \B{\xi}_p} - \dfrac{\partial \B{a}_p}{\partial \B{r}}\dfrac{\partial \B{r}}{\partial \B{\xi}_p} - \dfrac{\partial \B{a}_p}{\partial \dot{\B{r}}}\dfrac{\partial \dot{\B{r}}}{\partial \B{\xi}_p} \\
    \label{eq:loss_xitheta}
    \dfrac{\partial \mathcal{L}}{\partial \B{\xi}_{\vartheta}} &= \dfrac{\partial \ddot{\B{r}}}{\partial \B{\xi}_{\vartheta}} + \dfrac{\mu}{r^3}\dfrac{\partial \B{r}}{\partial \B{\xi}_{\vartheta}} - \dfrac{\partial \B{a}_p}{\partial \B{r}}\dfrac{\partial \B{r}}{\partial \B{\xi}_{\vartheta}} - \dfrac{\partial \B{a}_p}{\partial \dot{\B{r}}}\dfrac{\partial\dot{ \B{r}}}{\partial \B{\xi}_{\vartheta}} \\
    \label{eq:loss_xih}
    \dfrac{\partial \mathcal{L}}{\partial \B{\xi}_h} &= \dfrac{\partial \ddot{\B{r}}}{\partial \B{\xi}_h} - 3 \mu \dfrac{\B{r}}{r^4}\dfrac{\partial r}{\partial \B{\xi}_h}  + \dfrac{\mu}{r^3}\dfrac{\partial \B{r}}{\partial \B{\xi}_h} - \dfrac{\partial \B{a}_p}{\partial \B{r}}\dfrac{\partial \B{r}}{\partial \B{\xi}_h} - \dfrac{\partial \B{a}_p}{\partial \dot{\B{r}}}\dfrac{\partial \dot{\B{r}}}{\partial \B{\xi}_h}
    \end{alignat}
\end{subequations}
Note that the last two terms in Eqs.~\eqref{eq:loss_xir}-\eqref{eq:loss_xih} vary according to the perturbation type. The $J_2$ gravitational perturbation is given by Eq.~\ref{eq:acc_J2}, with partial derivatives:
\begin{align*}
    \mathcal{J}_{J2xx} &= \frac{3J_2\mu r_{eq}^2}{2r^7}\left[5x^2(1-5(z/r)^2)-r^2(1-5(z/r)^2)-10(xz/r)^2\right] \\
    \mathcal{J}_{J2xy} &= \frac{15J_2\mu r_{eq}^2xy}{2r^7}\left(1-7(z/r)^2\right) = \mathcal{J}_{J2yx}\\
    \mathcal{J}_{J2xz} &= \frac{15J_2\mu r_{eq}^2xz}{2r^7}\left(3-7(z/r)^2\right) = J_{J2zx}\\
    \mathcal{J}_{J2yy} &= \frac{3J_2\mu r_{eq}^2}{2r^7}\left[5y^2(1-5(z/r)^2)-r^2(1-5(z/r)^2)-10(yz/r)^2\right] \\
    \mathcal{J}_{J2yz} &= \frac{15J_2\mu r_{eq}^2yz}{2r^7}\left(3-7(z/r)^2\right) = \mathcal{J}_{J2zy}\\
    \mathcal{J}_{J2zz} &= \frac{3J_2\mu r_{eq}^2}{2r^7}\left[5z^2(3-5(z/r)^2)-r^2(3-5(z/r)^2)-10z^2(1-(z/r)^2)\right]
\end{align*}
The third-body perturbation is given by Eq.~\ref{eq:acc_3b}, and its partial derivatives are:
\begin{align*}
    \mathcal{J}_{3BPxx} &= \mu_{3b}\left(\frac{3x_{sc-3b}^2}{r_{sc-3b}^5}-\frac{1}{r_{sc-3b}^3}\right) & \mathcal{J}_{3BPxy} &= \mu_{3b}\left(\frac{3x_{sc-3b}y_{sc-3b}}{r_{sc-3b}^5}\right) = \mathcal{J}_{3BPyx}\\
    \mathcal{J}_{3BPyy} &= \mu_{3b}\left(\frac{3y_{sc-3b}^2}{r_{sc-3b}^5}-\frac{1}{r_{sc-3b}^3}\right) & \mathcal{J}_{3BPxz} &= \mu_{3b}\left(\frac{3x_{sc-3b}z_{sc-3b}}{r_{sc-3b}^5}\right) = \mathcal{J}_{3BPzx} \\
    \mathcal{J}_{3BPzz} &= \mu_{3b}\left(\frac{3z_{sc-3b}^2}{r_{sc-3b}^5}-\frac{1}{r_{sc-3b}^3}\right) & \mathcal{J}_{3BPzy} &= \mu_{3b}\left(\frac{3y_{sc-3b}z_{sc-3b}}{r_{sc-3b}^5}\right) = \mathcal{J}_{3BPyz}
\end{align*}
Finally, the SRP perturbation is given by Eq.~\ref{eq:acc_srp}, and its partial derivatives are:
\begin{align*}
    \mathcal{J}_{SRPxx} &= \frac{P_{SRP}A}{m}\bigg(\frac{2\rho_ax_s}{r_s^2}-\frac{2\rho_ax_s^3}{r_s^4}+\frac{4\rho_sx_s}{r_s^2}-\frac{4\rho_sx_s^3}{r_s^4} + \rho_d(\frac{x_s}{r_s^2}-\frac{2x_s^3}{r_s^4}+\frac{2}{3}(\frac{1}{r_s}-\frac{x_s^2}{r_s^3}))\bigg) \\
    \mathcal{J}_{SRPxy} &= \frac{P_{SRP}A}{m}\left[-\frac{2\rho_ax_s^2y_s}{r_s^4}-\frac{4\rho_sy_sx_s^2}{r_s^4}-\frac{\rho_dx_s^2y_s}{r_s^4}-\frac{\rho_dx_sy_s}{r_s^3}(\frac{2}{3}-\frac{x_s}{r_s})\right] \\
    \mathcal{J}_{SRPxz} &= \frac{P_{SRP}A}{m}\left[-\frac{2\rho_ax_s^2z_s}{r_s^4}-\frac{4\rho_sz_sx_s^2}{r_s^4}-\frac{\rho_dx_s^2z_s}{r_s^4}-\frac{\rho_dx_sz_s}{r_s^3}(\frac{2}{3}-\frac{x_s}{r_s})\right] \\
    \mathcal{J}_{SRPyx} &= \frac{P_{SRP}A}{m}\left[-\frac{2\rho_ay_sx_s^2}{r_s^4}-\frac{\rho_dy_sx_s^2}{r_s^4}+\frac{\rho_dy_s}{r_s}(\frac{1}{r_s}-\frac{x_s^2}{r_s^3})\right] \\
    \mathcal{J}_{SRPyy} &= \frac{P_{SRP}A}{m}\left[\frac{\rho_ax_s}{r_s^2}-\frac{2\rho_ax_sy_s^2}{r_s^4}+\frac{\rho_dx_s}{r_s}(\frac{1}{r_s}-\frac{y_s^2}{r_s^3})-\frac{\rho_dx_sy_s^2}{r_s^4}\right] \\
    \mathcal{J}_{SRPyz} &= \frac{P_{SRP}A}{m}\left[-\frac{2\rho_ax_sy_sz_s}{r_s^4}-\frac{2\rho_dx_sy_sz_s}{r_s^4}\right] = \mathcal{J}_{SRPzy}\\
    \mathcal{J}_{SRPzx} &= \frac{P_{SRP}A}{m}\left[\frac{\rho_az_s}{r_s^2}-\frac{2\rho_ax_s^2z_s}{r_s^4}-\frac{\rho_dx_s^2z_s}{r_s^4}+\frac{\rho_dz_s}{r_s}(\frac{1}{r_s}-\frac{x_s^2}{r_s^3})\right] \\
    \mathcal{J}_{SRPzz} &= \frac{P_{SRP}A}{m}\left[\frac{\rho_ax_s}{r_s^2}-\frac{2\rho_ax_sz_s^2}{r_s^4}+\frac{\rho_dx_s}{r_s}(\frac{1}{r_s}-\frac{z_s^2}{r_s^3})-\frac{\rho_dx_sz_s^2}{r_s^4}\right]
\end{align*}
Note that the partials of the SRP perturbation are simplified using the parameters chosen in this investigation.

\end{document}